\documentclass[a4paper,DIV=12]{scrartcl}

\setkomafont{disposition}{\normalfont\bfseries}

\usepackage{amsmath}
\usepackage{amsthm}
\usepackage{amssymb}
\usepackage{stmaryrd}
\usepackage[ruled,vlined]{algorithm2e}
\usepackage{mathrsfs}
\usepackage{enumerate}
\usepackage{mathtools}
\usepackage{xspace}
\usepackage{float}
\usepackage{url}
\usepackage{todonotes}
\usepackage{multicol}
\usepackage{listofitems}

\usetikzlibrary{automata,positioning,calc}

\newcommand{\FO}{\textnormal{FO}}
\usepackage[font=sf,width=.9\linewidth]{caption}

\newcounter{continueListCounter}

\newcommand{\Si}{\mathbf{Si}}

\newcommand{\DA}{\mathbf{DA}}

\newcommand{\relmor}[6]{\begin{tikzcd} $C$ \end{tikzcd}}

\newcommand{\alp}{\mathsf{alph}}
\newcommand{\faktor}[2]{#1 / {#2}}
\newcommand{\intp}[1]{\llbracket #1 \rrbracket}
\newcommand{\genby}[1]{\langle #1 \rangle}

\newcommand{\floor}[1]{\lfloor #1 \rfloor}

\newcommand{\malcev}{\mathbin{
  \mathchoice
    {\mbox{\normalsize\textup{\textcircled{\scriptsize M}}}}
    {\mbox{\normalsize\textup{\textcircled{\scriptsize M}}}}
    {\mbox{\scriptsize\textup{\textcircled{\tiny M}}}}
    {\mbox{\tiny\textup{\textcircled{\fontsize{3.5}{3.5}\selectfont M}}}}
  }
}

\newcommand{\genmalcev}[1]{
  \setsepchar{_}%
  \readlist\mymat{#1}%
  \ifnum\mymatlen>1 \left(#1\right)_{\mathbf{KD}} \else #1_{\mathbf{KD}} \fi}
\newcommand{\startof}[1]{\ensuremath{\rhd #1}}

\newcommand{\greenfont}[1] {\ensuremath{\mathcal{#1}}}

\newcommand{\greenR} {\greenfont{R}\xspace}
\newcommand{\Req}    {\mathrel{\greenR}}
\newcommand{\Rleq}   {\mathrel{\leq_\greenR}}

\newcommand{\greenL} {\greenfont{L}\xspace}
\newcommand{\Leq}    {\mathrel{\greenL}}
\newcommand{\Lleq}   {\mathrel{\leq_\greenL}}

\newcommand{\greenJ} {\greenfont{J}\xspace}
\newcommand{\Jeq}    {\mathrel{\greenJ}}
\newcommand{\Jleq}   {\mathrel{\leq_\greenJ}}

\newcommand{\itref}[1]{\textit{(\ref{#1})}}

\title{Deciding FO$\mathbf{^2}$ Alternation for Automata over Finite and Infinite Words}

\author{Viktor Henriksson\hspace*{1pt}$^1$ and Manfred Kuf\-leitner\hspace*{2pt}$^2$}
\date{\normalsize$^1$ Loughborough University, Loughborough, UK \\
\texttt{b.v.d.henriksson@lboro.ac.uk} \\[3mm]
$^2$ University of Stuttgart, Stuttgart, Germany \\
\texttt{kufleitner@fmi.uni-stuttgart.de}}

\newcommand{\semleq}[1]{\leq_{#1}}
\newcommand{\semgeq}[1]{\geq_{#1}}
\newcommand{\semeq}[1]{\equiv_{#1}}
\newcommand{\NL}{\ensuremath{\mathbf{NL}}\xspace}
\newcommand{\Pspace}{\ensuremath{\mathbf{PSPACE}}\xspace}
\newcommand{\coNP}{\ensuremath{\mathbf{coNP}}\xspace}
\DeclareMathOperator{\im}{im}

\newtheorem{definition}{Definition}
\newtheorem{proposition}{Proposition}
\newtheorem{theorem}{Theorem}
\newtheorem{corollary}{Corollary}
\newtheorem{lemma}{Lemma}
\newtheorem*{claim}{Claim}

\usepackage{refcount}

\newtheorem{cusprp}{Proposition}

\newtheorem{custhm}{Theorem}

\newtheorem{cuscor}{Corollary}

\begin{document}

\maketitle

\begin{abstract}
	We consider two-variable first-order logic $\FO^2$ and its quantifier alternation hierarchies over both finite and infinite words. Our main results are forbidden patterns for deterministic automata (finite words) and for Carton-Michel automata (infinite words). In order to give concise patterns, we allow the use of subwords on paths in finite graphs. This concept is formalized as subword-patterns. For certain types of subword-patterns there exists a non-deterministic logspace algorithm to decide their presence or absence in a given automaton.
	In particular, this leads to $\NL$ algorithms for deciding the levels of the $\FO^2$ quantifier alternation hierarchies. This applies to both full and half levels, each over finite and infinite words. Moreover, we show that these problems are $\NL$-hard and, hence, $\NL$-complete.
\end{abstract}

\section{Introduction}

Many interesting varieties of finite monoids can be defined by a finite set of identities of $\omega$-terms. By Eilenberg's Variety Theorem~\cite{eilenberg1974}, every variety of finite monoids corresponds to a unique variety for regular languages. In particular, identities of $\omega$-terms can be used for describing classes of regular languages. If $L \subseteq A^*$ is given by a homomorphism $\varphi : A^* \to M$ to a finite monoid together with an accepting set $P \subseteq M$ such that $L = \varphi^{-1}(P)$, then one can check in nondeterministic logarithmic space \NL whether $M$ satisfies a fixed identity of $\omega$-terms; see e.g.~\cite[Theorem~2.19]{straubingweil2015TR} or~\cite{fleischer2018TR}. If $L$ is given by a (deterministic or nondeterministic) finite automaton, then this algorithms yields a \Pspace-algorithm for deciding whether $L$ satisfies the identity (by applying the \NL algorithm to the transition monoid of the automaton; in the case of nondeterministic automata, this monoid can be represented by Boolean matrices). Since universality of nondeterministic automata is \Pspace-complete~\cite{kozen1977}, there is no hope for more efficient algorithms if $L$ is given by a nondeterministic automaton.

The star-free languages can be defined by a very short identity of $\omega$-terms~\cite{shutzenberger1976sf}. In 1985, Stern showed that deciding whether a given deterministic automaton accepts a star-free language is \coNP-hard, leaving open whether it was in fact \Pspace-complete~\cite{Stern85ic}. This was later given an affirmative answer by Cho and Huynh~\cite{ChoHuynh91tcs}.
For other important varieties, the situation is very different. In the same paper, Stern gave polynomial time algorithms for deciding membership of the $\mathcal{J}$-trivial (also referred to as piecewise testable) languages and languages of dot-depth one~\cite{Stern85ic} when the languages are given by deterministic finite automata. The exact complexity for these problems was again given by Cho and Huynh, showing that they are \NL-complete~\cite{ChoHuynh91tcs}.

Forbidden patterns are a common approach for efficiently solving the membership problem. Stern's polynomial time algorithms build on pattern characterizations~\cite{Stern85tcs}. Characterizations of $\mathcal{R}$ and $\mathcal{L}$-trivial languages using forbidden patterns were given by Cohen et al.~\cite{CohenPerrinPin93jcss}, and Schmitz et al. used the approach for characterizing the first levels of the Straubing-Th\'erien hierarchy~\cite{GlasserSchmitz08tocs,SchmitzWagner98Arxiv}.

The pattern approach usually relies on the DFA of a language. Since deterministic B\"uchi automata cannot express all $\omega$-regular languages, this has inhibited the adaptation of the pattern approach in the study of $\omega$-regular languages. In 2003, Carton and Michel introduced a type of automata~\cite{CartonMichel03tcs}, (originally called complete unambigous B\"uchi automata, but nowadays known as Carton-Michel automata) which they showed to be expressively complete for $\omega$-regular languages. These automata associate every word to a unique path, making it an ideal candidate for using patterns in the context of $\omega$-regular languages. Preugschat and Wilke~\cite{preugschatwilke2013lmcs} pioneered this approach by giving characterizations of fragments of temporal logic relying partly on patterns. Their method involved separating the finite behaviour of the language from the infinite behaviour; the finite behaviour was then characterized using patterns, while the infinite behaviour was characterized using conditions on loop languages.

The variety of languages definable in $\FO^2$, i.e., first order logic with only two variables, is well studied. Th\'erien and Wilke~\cite{TherienWilke1998stoc} showed that this variety was the collection of languages whose syntactic monoid was in $\DA$. In particular, this established an equivalence between $\FO^2$ and 
$\Sigma_2 \cap \Pi_2$ over finite words.

One can consider the quantifier alternation hierarchy inside $\FO^2$. Due to the restriction on the number of variables, one needs to consider parse trees rather than translating formulae into prenex normal form. 
Over finite words, Weis and Immerman gave a combinatorial characterization of the join levels of this hierarchy~\cite{WeisImmerman2009lmcs}; algebraic characterizations were given by Weil and the second author~\cite{kufleitnerweil2012csl} and independently by Krebs and Straubing~\cite{KrebsStraubing2017tocl}. The half-levels were characterized by Fleischer, Kuf\-leitner and Lauser~\cite{FleischerKL17tocs}.

For $\omega$-regular languages, algebraic characterizations often utilize Arnold's congruence. However, not every interesting class of languages can be characterized directly using this congruence; see e.g.~\cite{PerrinPin04}. On the other hand, combining algebraic properties with topology has proven a fruitful alternative in some cases where algebra alone is not enough; see e.g.~\cite{diekertkufleitner2011tocs,KallasEtAl2011stacs,KufleitnerWalter2018tocs}. In particular, this approach was used in yet unpublished work by Boussidan and the second author for the characterization of the join levels of the alternation hierarchies, and by the authors for the characterization of the half-levels~\cite{HenrikssonKufleitner20Arxiv}.

This article is outlined as follows. In Section \ref{sec:prelim}, we give brief introductions to the three main areas of this article, \emph{formal languages}, \emph{monoids} and \emph{logic}. We devote Section \ref{sec:subwordpatterns} to the development of  a formalism for \emph{subword-patterns}: patterns where we can not only use identical words as labels of different paths, but also subwords. Patterns taking subwords into account were used, e.g.\, in~\cite{SchmitzWagner98Arxiv}. Our formalism is a variation of that of Kl\'ima and Pol\'ak~\cite{KlimaPolak20jalc}, but considering automata instead of ordered semiautomata. For DFAs, this difference is superficial since the relevant semi-DFA can be obtained via minimization. Minimizing a Carton-Michel automaton (based on the reverse deterministic transition relation) does not necessarily produce a Carton-Michel automaton.
For patterns which do not take final states into account, such as those used in~\cite{preugschatwilke2013lmcs}, this is not a problem. However, this contribution contains patterns for which it matters.

In Section \ref{sec:Hierarchies}, we use the mentioned formalism to give DFA patterns for the algebraic varieties used in the characterizations of the quantifier alternation hierarchies inside $\FO^2$. Section \ref{sec:reverse} contains an interlude in which we give some technical details on how patterns for DFAs can be transfered to patterns for reverse-DFA. This is crucial for our treatment of patterns for Carton-Michel automata in Sections \ref{sec:PatternsForCartonMichelAutomata} and \ref{sec:InfBehaviour}.

We split the problem of deciding membership for Carton-Michel automata into two parts, dealing with the finite and infinite behaviour respectively. The finite behaviour, as well as the formalization of this split, is dealt with in Section \ref{sec:PatternsForCartonMichelAutomata}. We deal with the so-called fin-syntactic monoid, and show that its membership of some variety can be characterized by the same pattern as in the finite case.

In dealing with the infinite behaviour in Section \ref{sec:InfBehaviour} we consider two behaviours. First, we consider the inf-syntactic monoid, show that it is enough to show its membership in $\DA$ and give a pattern for deciding this. Next, we consider topology and give patterns for open and closeness in the Cantor and alphabetic topology.

Finally, Section \ref{sec:Complexity} deals with complexity. We show that for any subword-pattern which has stable superwords, presence in a given DFA or Carton-Michel automata is in $\NL$. This in particular shows that membership in $\FO^2_m$ for these inputs is in $\NL$ for all $m$. We also show $\NL$-hardness, showing that these problems are $\NL$-complete.

\section{Preliminaries}
\label{sec:prelim}


\subsection{Languages and Automata}	
\label{sec:LangAndAutomata}

For an alphabet $A$ we denote by $A^*$ the set of finite words over $A$ and by $A^{\omega}$ the set of infinite words over $A$. A subset $L \subseteq A^*$ or $L \subseteq A^{\omega}$ is a \emph{language}.
If $u = a_1 \dots a_n \in A^*$ is a word, then $\overline{u} = a_n \dots a_1$, and if $L \subseteq A^*$ is a language, then $\overline{L} = \left\{ \overline{u} \mid u \in L \right\}$. 
The \emph{alphabet} of $u \in A^*$, $\alp(u)$, is the set of letters which occurs in $u$, and the \emph{imaginary alphabet} of $\alpha \in A^{\omega}$, $\im(\alpha)$, is the set of letters $a$ which occurs on infinitely many positions of $\alpha$.
For a word $u \in A^*$, we denote by $u^{\omega} = uuu\cdots$ the infinite iteration of $u$. 

A \emph{language variety} is a system $\mathcal{V}$ which to each alphabet $A$ associate a set of languages $\mathcal{V}_{A} \subseteq 2^{A^*}$ in such a way that:
\begin{enumerate}[\itshape(i)]
	\item $L,L' \in \mathcal{V}_A$ implies $L \cap L' \in \mathcal{V}_A$, $L \cup L' \in \mathcal{V}_A$,\label{aaa:variety}
	\item $L \in \mathcal{V}_A$ implies $A^* \setminus L \in \mathcal{V}_A$,\label{bbb:variety}
	\item $L \in \mathcal{V}_A$ implies $u^{-1}Lv^{-1} = \left\{ w \in A^* \mid uwv \in L \right\} \in \mathcal{V}_A$.\label{ccc:variety} 
	\item for every map $h: B^* \to A^*$, $L \in \mathcal{V}_A$ implies $h^{-1}(L) \in \mathcal{V}_B$.\label{ddd:variety}
\end{enumerate}
A language of the form $u^{-1}Lv^{-1}$ is called a \emph{residual}. In particular, if $v = \varepsilon$, then it is a \emph{left-residual} and if $u = \varepsilon$ a \emph{right-residual}.
If conditions \itref{aaa:variety}, \itref{ccc:variety} and \itref{ddd:variety} but not necessarily \itref{bbb:variety} is satisfied, we call it a \emph{positive variety}.

A \emph{deterministic finite automaton (DFA)} is a tuple $\mathcal{A} = \left( Q, A, \cdot,i, F \right)$ where:
\begin{itemize}
	\item $Q$ is a finite set of \emph{states},
	\item $A$ is an alphabet,
	\item $\cdot: Q \times A \to Q$ is a \emph{transition function},
	\item $i \in Q$ is an \emph{initial state},
	\item $F \subseteq Q$ is a set of \emph{final states}.
\end{itemize}
A \emph{semi}-DFA is a DFA without the initial state $i$ and the final states $F$, and a semi-DFA is \emph{partial} if $\cdot$ is a partial function.

 We can extend $\cdot$ to a function $Q \times A^* \to Q$ by $j \cdot (a_1 \dots a_n) = ((j \cdot a_1) \cdot \ldots) \cdot a_n$. For $u \in A^*$, we say that $\mathcal{A}$ \emph{accepts} $u$ if $i \cdot u \in F$. We define
\begin{equation*}
	L(\mathcal{A}) = \left\{ u \in A^* \mid \text{$\mathcal{A}$ accepts $u$} \right\}.
\end{equation*}
Then $\mathcal{A}$ \emph{accepts} $L$ if $L = L(\mathcal{A})$.

Since the number of states in a (partial semi-)DFA $\mathcal{A}$ is finite, there exists a number $\eta_{\mathcal{A}}$ such that $j \cdot u^{\eta_{\mathcal{A}}} u^{\eta_{\mathcal{A}}} = j \cdot u^{\eta_{\mathcal{A}}}$ for all $u \in A^*$. When $\mathcal{A}$ is clear from context, we simply write $\eta$.

Given two partial semi-DFAs $\mathcal{A} = (Q,A,\cdot)$ and $\mathcal{A}' = (Q',A,\cdot')$, $f: Q \to Q'$ is a \emph{homomorphism of partial semiautomata} if $f(j \cdot a) = f(j) \cdot' a$ for all $j \in Q$, $a \in A$ such that $j \cdot a$ is defined. A homomorphism of partial semiautomata is a homomorphism of DFAs if the partial semiautomata are also DFAs, say $\mathcal{A} = (Q,A,\cdot,i,F)$ and $\mathcal{A}' = (Q',A,\cdot',i',F')$, and $f(i) = i'$ and $f^{-1}(F') = F$.

A \emph{reverse DFA} is a tuple $\mathcal{A} = (Q,A,\circ,I,f)$ where $Q$ and $A$ are as in a DFA, $\circ: Q \times A \to Q$ is a reverse transition function, $I \subseteq Q$ is a set of initial states, and $f \in Q$ is a final state. Note that there is no formal difference between a transition and a reverse transition function. The difference lies in the interpretation and the extension to $A^*$; we write $a \circ j$ for the value at $(j,a)$ and we define $a_1 \dots a_n \circ j = a_1 \circ ( \dots \circ (a_n \circ j))$. Thus the function is applied in the reverse order, starting with $a_n$. If $u \circ f \in I$, then $\mathcal{A}$ accepts $u$.
If $\mathcal{A} = (Q,A,\cdot,i,F)$ is a DFA, then $\overline{\mathcal{A}} = \left( Q,A,\cdot,F,i \right)$ is a reverse DFA and accepts $\overline{L(\mathcal{A})}$. Conversely, if $\mathcal{A} = \left( Q,A,\circ,I,f \right)$ is a reverse DFA, then $\overline{\mathcal{A}} = \left( Q,A,\circ,f,I \right)$. 

\paragraph{Carton-Michel Automata}

We introduce \emph{Carton-Michel automata}, a particular type of B\"uchi automata.
Let $\mathcal{A} = (Q,A, \circ, I, F)$ be a B\"uchi automaton. A \emph{run} of $\mathcal{A}$ is an infinite path in $\mathcal{A}$. Each such run is \emph{labeled} by an infinite word by reading the letters corresponding to each edge of the path. A run is \emph{final} if it visits a final state infinitely often. The run is \emph{accepting} if it is final and starts at an initial state. A word is \emph{accepted} by $\mathcal{A}$ if it labels some accepting run, and the language \emph{accepted} by $\mathcal{A}$, denoted $L(\mathcal{A})$, is the collection of all such words.

A \emph{Carton-Michel automaton} $\mathcal{A}$ is a B\"uchi automaton where every infinite word has a unique final run. In particular, this means that for each word $\alpha \in A^{\infty}$, we can associate a state in $\mathcal{A}$. We denote this state $\startof \alpha$. 
The following theorem gives one of the key points of Carton-Michel automata.

\begin{theorem}[Carton and Michel \cite{CartonMichel03tcs}]
	Every $\omega$-regular language is accepted by some Carton-Michel automata.
\end{theorem}
A subautomaton  $\mathcal{B}$ of $\mathcal{A}$ is \emph{trim} if it is a Carton-Michel automata and for every state $j \in \mathcal{B}$, there exists $\alpha_j \in A^*$ such that $j = \startof \alpha_j$. 
As noted by Carton and Michel, a trim Carton-Michel automata is reverse deterministic \cite{CartonMichel03tcs}.\footnote{Note that the Carton-Michel automata used by Preugschat and Wilke for their pattern approach~\cite{preugschatwilke2013lmcs} have a slight technical difference, where the automata are assumed to be reverse-deterministic. These definitions coincide on all trim Carton-Michel automata.}

\paragraph{Topology}

A set $\mathcal{T} \subseteq 2^{A^*}$ is a \emph{topology} if $\emptyset, A^* \in \mathcal{T}$ and $\mathcal{T}$ is closed under finite intersections and arbitrary unions. A language $L \subseteq A^*$ is \emph{open} if $L \in \mathcal{T}$, and \emph{closed} if its complement is in $\mathcal{T}$. A set $\mathcal{B} \subseteq 2^{A^*}$ is a \emph{base} for a topology if $A^* = \bigcup_{L \in \mathcal{B}} L$ and if for all $L_1, L_2 \in \mathcal{B}$, there exists $\mathcal{B}' \subseteq \mathcal{B}$ such that $L_1 \cap L_2 = \bigcup_{K \in \mathcal{B}'} K$. The sets of unions of elements in $\mathcal{B}$ is a topology, the topology \emph{generated} by $\mathcal{B}$.

For a DFA $\mathcal{A} = \left( Q, A, \cdot, i, F  \right)$, we say that $j \semleq{\mathcal{A}} k$ if $j \cdot u \in F$ implies $k \cdot u \in F$ for all $u \in A^*$. We say that $i \semeq{\mathcal{A}} j$ if $i \semleq{\mathcal{A}} j$ and $j \semleq{\mathcal{A}} i$. We use the same notation for reverse DFAs and Carton-Michel automata; we say $j \semleq{\mathcal{A}} k$ if $u \circ j \in I$ implies $u \circ k \in I$, and $j \semeq{\mathcal{A}} k$ if $j \semleq{\mathcal{A}} k$ and $k \semleq{\mathcal{A}} j$.

The \emph{Cantor topology} $\mathcal{O}_{cantor}$ is the topology generated by the base $\left\{ uA^{\omega} \right\}_{u\in A^*}$, and the \emph{alphabetic topology} $\mathcal{O}_{alph}$ is the topology generated by $\left\{ uB^{\omega} \right\}_{u\in A^*, B \subseteq A}$. We denote by $\mathbb{B}(\mathcal{T})$ the Boolean closure of a topology.

\subsection{Monoids, Varieties and Recognition}


Let $M$ be a monoid generated by a set $A$. The \emph{Cayley-graph} of $M$ is the semiautomata $\mathcal{G} = (M,A,\cdot)$ where $\cdot$ is defined by $m \cdot a = ma$ for all $m \in M$, $a \in A$. The Cayley-graph has a root (and natural initial state) given by the unit of $M$.

Every monoid $M$ has a number $\omega_M$ such that $s^{\omega_M}s^{\omega_M} = s^{\omega_M}$ for all $s \in M$. We call $\omega_M$ the \emph{idempotent power}. If $M$ is clear from context, we only write $\omega$. An element $e \in M$ is \emph{idempotent} if $e = e^{\omega_M}$. A pair $(s,e)$ is \emph{linked} if $e$ is idempotent and $se = s$.

Given a binary relation $\preceq$ on a monoid $M$, the relation is \emph{stable} if $s \preceq t$ implies $psq \preceq ptq$. A conjugacy is a relation which is reflexive and stable. Every stable relation $\preceq$ induces a conjugacy $\sim$ by $s \sim t$ if $s \preceq t$ and $t \preceq s$.
A monoid with a stable partial order is an \emph{ordered monoid}.  A homomorphism $h: M \to N$ between ordered monoids is \emph{monotone} if $s \leq t$ implies $h(s) \leq h(t)$. Note that ordered monoids generalizes monoids, since any monoid can use the equality relation as an order.
If $\preceq$ is stable on $M$, then $\faktor{M}{\preceq}$ is the monoid consisting of the equivalence classes of the induced conjugacy, and the order induced by $\preceq$.

An important tool in the study of finite monoids are the \emph{Green's relations}. We introduce the relations \greenR, \greenL and \greenJ. Let $s, t \in M$, then
\begin{itemize}
	\item $s \Rleq t$ if $sM \subseteq tM$,
	\item $s \Lleq t$ if $Ms \subseteq Mt$,
	\item $s \Jleq t$ if $MsM \subseteq MtM$,
\end{itemize}
and $s \Req t$ if $s \Rleq t$ and $t \Rleq s$. The relations \greenL and \greenJ are defined analogously from $\Lleq$ and $\Jleq$ respectively.

\paragraph{Varieties}

The ordered monoid $N$ \emph{divides} $M$, if there exists a submonoid $M' \subseteq M$ and a surjective monotone homomorphism $h: M' \to N$. A class $\mathbf{V}$ of finite ordered monoids is a \emph{variety of ordered monoids} if $M, N \in \mathbf{V}$ implies $M \times N \in \mathbf{V}$ and $M' \in \mathbf{V}$ for all $M'$ which divides $M$. A variety of unordered monoids is defined analogously, but for unordered monoids and homomorphisms which are not necessarily monotone.
Unless specified otherwise, we use the following notation: suppose $\mathbf{V}$ is a variety of (ordered) monoids; then $\mathcal{V}$ is the (positive) variety of languages whose syntactic monoids are in $\mathbf{V}$.

Let $\Omega$ be a set of variables. The set of $\omega$-terms, $T(\Omega)$, over $\Omega$ is defined inductively: $\Omega \cup \left\{ 1 \right\} \subseteq T(\Omega)$ and if $x, y \in T(\Omega)$ then $xy \in T(\Omega)$ and $x^{\omega} \in T(\Omega)$. Here $x^{\omega}$ is a formal symbol, not related to the infinite concatenation, and not strictly the same as the $\omega$ denoting the idempotent power. The meaning of the symbol will be clear from context. 
An \emph{interpretation} is a function $I : \Omega \to M$. Any such function can be extended to a function $I: T(\Omega) \to M$ by setting $I(ts) = I(t)I(s)$ and $I(t^{\omega}) = (I(t))^{\omega_M}$ for all $t,s \in T(\Omega)$. If $s,t \in T(\Omega)$, then $M$ \emph{satisfy} $s \leq t$ if $I(s) \leq I(t)$ for all interpretations. Satisfiability of $s = t$ is defined analogously.
We define $\intp{s \leq t}$ to be the collection of monoids which satisfy $s \leq t$, and we define $\intp{s = t}$ analogously. Any collection defined in this way is a positive variety (and in the latter case also a nonpositive variety).

The following varieties are of particular importance throughout this contribution:
\begin{itemize}
	\item $\DA = \intp{((yx)^{\omega} y (yx)^{\omega} = (xy)^{\omega}}$,
	\item 
	 $\mathbf{R} = \intp{(yx)^{\omega} y = (yx)^{\omega}}$,
	 $\mathbf{L} = \intp{y(xy)^{\omega} = (xy)^{\omega}}$,
	\item $\mathbf{J}_1 = \intp{z^2 = z, xy = yx}$,
	$\mathbf{J}^+ = \intp{1 \leq z}$
\end{itemize}
We record the following well-known property of $\DA$ (see e.g.~\cite{diekertgastinkufleitner2008ijfcs}).

\begin{lemma}\label{lem:DAproperty}
	Let $M \in \DA$ and let $\mu: A^* \to M$ be a homomorphism. Then $\alp(u) \subseteq \alp(v)$ implies $\mu(v)^{\omega}\mu(u)\mu(v)^{\omega} = \mu(v)^{\omega}$ for all $u, v \in A^*$.
\end{lemma}

One way to generate new varieties from known ones is by using the Malcev product. Generally, Malcev products are defined using relational morphism. However, for the two semigroup varieties $\mathbf{K}$ and $\mathbf{D}$, a more direct approach using the relations $\sim_{\mathbf{K}}$ and $\sim_{\mathbf{D}}$ is sufficient. This approach was refined in~\cite{HenrikssonKufleitner20Arxiv} to define a chain of ordered monoids. Let $s, t \in M$, then:
\begin{itemize}
	\item $s \sim_{\mathbf{K}} t$ if for all idempotent elements $e$, we have $es,et <_{\mathcal{J}} e$ or $es = et$,
	\item $s \sim_{\mathbf{D}} t$ if for all idempotent elements $f$, we have $sf,tf <_{\mathcal{J}} f$ or $sf = tf$,
	\item $s \preceq_{\mathbf{KD}} t$ if for all $p,q \in M$:
	$p \Req ptq$ implies $p \Req psq$, $ptq \Leq q$ implies $psq \Leq q$, and $p \Req pt \,\wedge\, tq \Leq q$ implies $psq \leq ptq$.
	
\end{itemize}

Given a variety $\mathbf{V}$, we say that $M \in \mathbf{K}  \malcev \mathbf{V}$ if $\faktor{M}{\sim_{\mathbf{K}}} \in \mathbf{V}$, $M \in \mathbf{D}  \malcev \mathbf{V}$ if $\faktor{M}{\sim_{\mathbf{D}}} \in \mathbf{V}$ and $M \in \genmalcev{\mathbf{V}}$ if $\faktor{M}{\preceq_{\mathbf{KD}}} \in \mathbf{V}$.
Let:
\begin{itemize}
	\item $\mathbf{R}_1 = \mathbf{L}_1 = \mathbf{R} \cap \mathbf{L}$, $\mathbf{R}_{m+1} = \mathbf{K} \malcev \mathbf{L}_m$, $\mathbf{L}_{m+1} = \mathbf{D} \malcev \mathbf{R}_m$,
	\item $\Si_1 = \mathbf{J}^+$, $\Si_{m+1} = \genmalcev{\Si_m}$.
\end{itemize}
It is well known that $\mathbf{R}_2 = \mathbf{R}$, $\mathbf{L}_2 = \mathbf{L}$ (see e.g.~\cite{Pin86}).

\paragraph{Syntactic Monoids}

Given a language $L \subseteq A^*$, we define $u \leq_L v$ for $u,v \in A^*$ if for all $p,q \in A^{*}$, $puq \in L \Rightarrow pvq \in L$. The \emph{syntactic morphism} of $L$ is the natural projection $\mu: A^* \to \faktor{A^*}{\leq_{L}}$, and $\faktor{A^*}{\leq_L}$ is the \emph{syntactic monoid}. Similarly, if $L \subseteq A^{\omega}$, we define $u \leq_L v$ if for all $p,q,w \in A^*$,
\begin{align*}
	puqw^{\omega} \in L &\Rightarrow pvqw^{\omega} \in L &\text{and}&&
	p(uw)^{\omega} \in L &\Rightarrow p(vw)^{\omega} \in L.
\end{align*}
The syntactic morphism and monoid are analogous to the finite case. For a language $L$ with a syntactic morphism $\mu: A^* \to M$, we say that the morphism $\nu: A^* \to N$ \emph{recognizes} $L$ if there exists a monotone homomorphism $h: N \to M$ such that $\mu = h \circ \nu$. If $\nu: A^* \to M$ is clear from context and $s \in M$, then we use the notation $[s] = \nu^{-1}(s)$.


\subsection{Fragments of Logic}

\begin{table}
	\caption{Decidability criteria for a language $L$ with syntactic monoid $M$}\label{tbl:MonoidCriteria}
	\centering
	{
	
	\smallskip
	
	\begin{tabular}{|c|c|c|}
		\hline& Finite Words & Infinite Words
		\\\hline $\Sigma^2_1$ & $M \in \Si_1$ & 
		\begin{tabular}{c}
			$M \in \Si_1$ \\ $L \in \mathcal{O}_{cantor}$ 
		\end{tabular}
		\\\hline $\FO^2_1$ & $M \in \mathbf{R} \cap \mathbf{L}$ & 
		\begin{tabular}{c}
			$M \in \mathbf{R} \cap \mathbf{L}$ \\ $L \in \mathbb{B}(\mathcal{O}_{cantor})$ 
		\end{tabular}
		\\\hline $\Sigma^2_2$ & $M \in \Si_2$ & 
		\begin{tabular}{c}
			$M \in \Si_2$ \\ $L \in \mathcal{O}_{alph}$
		\end{tabular}
		\\\hline $\FO^2_m$, $m \geq 2$ & \multicolumn{2}{c|}{$M \in \mathbf{R}_{m+1} \cap \mathbf{L}_{m+1}$} 
		\\\hline $\Sigma^2_m$, $m \geq 3$ & \multicolumn{2}{c|}{$M \in \Si_{m}$} 
		\\\hline
	\end{tabular}}
\end{table}

Let $A = \left\{ a_1,\dots,a_n \right\}$ be an alphabet. We consider the fragment $\FO^2$ of first order logic over the signature $(\leq,a_1,\dots,a_n)$ where we only allow the use (and reuse) of two different variables. This fragment can be restricted further, by considering the number of allowed alternations. 
Consider the syntax
\begin{align*}
	\varphi_{0} &::= \top \mid \bot \mid \lambda(x) = a \mid \lambda(y) = a \mid x=y \mid x<y \mid y < x \mid \neg \varphi_{0} \mid \varphi_0 \vee \varphi_0 \mid \varphi_0 \wedge \varphi_0\\
	\varphi_{m} &::= \varphi_{m-1} \mid \neg \varphi_{m-1} \mid \varphi_{m} \vee \varphi_{m} \mid \varphi_{m} \wedge \varphi_{m} \mid \exists x \varphi_{m} \mid \exists y \varphi_m
\end{align*}
where $a \in A$, and $x$ and $y$ are (fixed) variables. The fragment $\Sigma^2_{m}$ consists of all formulae $\varphi_m$, the fragment $\Pi^2_m$ of all negations of formulae in $\Sigma^2_m$ and the fragment $\FO^2_m$ of the Boolean combinations of formulae in $\Sigma^2_m$.

Each of these logical fragments defines a language variety. These varieties have decidability characterizations for both finite words \cite{kufleitnerweil2012csl,FleischerKL17tocs} and infinite words \cite{BoussidanKufleitner18Unpub,HenrikssonKufleitner20Arxiv}. These criteria are presented in Table \ref{tbl:MonoidCriteria}.

\section{Subword-Patterns}
\label{sec:subwordpatterns}

In this section, we introduce subword-patterns. Our formalism is inspired by that of Kl\'ima and Pol\'ak~\cite{KlimaPolak20jalc}, with two main differences; we work with DFAs instead of ordered semi-DFAs, and we allow our patterns to take subwords into account.

In $\DA$, there is semantic equivalence between being a subword of and a factor of sufficiently long words. Thus, the patterns introduced in Section \ref{sec:Hierarchies} can be rewritten to equivalent patterns which do not rely on subwords. 
However, the patterns obtained in this way are less readable than their equivalent subword-patterns, arguably giving less insight into the actual behaviour of the varieties in consideration.

For the definition of subword-patterns, we rely on homomorphism of semi-DFAs. The following definition is standard, and gives a way to define homomorphisms between semi-DFAs which originally had different alphabets.

\begin{definition}
	Let $\mathcal{A} = \left( Q,A,\cdot \right)$ be a semi-DFA, and let $h: B^* \to A^*$ be a homomorphism. The \emph{$h$-renaming} of $\mathcal{A}$ is the semi-DFA $\mathcal{A}^h = \left( Q,B,\cdot^h \right)$ where $i \cdot^h b = i \cdot h(b)$.
\end{definition}

We give the formal definition of a subword-pattern. Intuitively, we can think of the edges of the pattern as paths in a given automata and the relation $\preceq$ as being the subword relation on the words labeling these paths.

\begin{definition}\label{def:subwordpatterns}
	Let $X$ be a set with a partial order $\preceq$. 
	A \emph{type 1 subword-pattern} $\mathcal{P} = (\mathcal{S},j \neq k)$ or \emph{type 2 subword-pattern} $\mathcal{P} = (\mathcal{S},j \not \leq k)$ consists of a finite partial semiautomaton $\mathcal{S} = \left( V, X, \cdot \right)$ and two states $j,k \in V$. If $\mathcal{P} = \left( \mathcal{S}, j \neq k \right)$, we say that $\mathcal{P}$ is \emph{present} in an automaton $\mathcal{A}$ if there exists a homomorphism $h: X^* \to A^*$ where $x \preceq y$ implies that $h(x)$ is a subword of $h(y)$ and a semiautomata homomorphism $g: \mathcal{S} \to \mathcal{A}^h$ such that $g(j) \not \semeq{\mathcal{A}} g(k)$ and for all $\ell \in V$, the state $g(\ell)$ is reachable from the initial state of $\mathcal{A}$.  Analogously, we say that $\mathcal{P} = \left( \mathcal{S}, j \not \leq k \right)$ is present if there exist $h$ and $g$ such that $g(j) \not \semleq{\mathcal{A}} g(k)$. Since the type of the pattern is clear from the notation, we usually do not reference its type.

	We say that a pattern is \emph{rooted} if there is some state $r \in V$ such that every $i$ satisfies $i = r \cdot x$ for some $x \in X^*$. Finally, two patterns $\mathcal{P}_1$, $\mathcal{P}_2$ are \emph{equivalent} if for all $\mathcal{A}$, the pattern $\mathcal{P}_1$ is present in $\mathcal{A}$ if and only if $\mathcal{P}_2$ is.
\end{definition}

Let us consider the following example. Let $X = \left\{ y,A_y \right\}$ with $A_y \preceq y$ and let $\mathcal{P}_{\DA} = \left( \mathcal{S},j \neq k \right)$ where $\mathcal{S}$ is
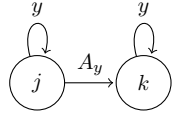
\begin{equation*}
	\begin{tikzpicture}[shorten >=1pt,node distance=1.4cm,on grid,auto,initial text = {},every node/.style={scale = 0.75}] 
		\node[state] (q_0) {$j$}; 
		\node[state] (q_2) [right=of q_0] {$k$}; 
	    \path[->] (q_0) edge[loop above]  node {$y$} (q_0);
	    \path[->] (q_0) edge  node {$A_y$} (q_2);
	    \path[->] (q_2) edge[loop above]  node {$y$} (q_2);
	\end{tikzpicture}
\end{equation*}
This pattern is present in an automata $\mathcal{A}$, if there are two cycles starting at different states, but labeled by the same word, as well as a path between them labeled by a word which is a subword of the aforementioned one. The following proposition shows that this pattern characterises having syntactic monoid in $\DA$.

\begin{proposition}
	Let $\mathcal{A}$ be an automata, and let $M$ be the syntactic monoid of $L(\mathcal{A})$. Then $M \in \DA$ if and only if $\mathcal{P}_{\DA}$ is not present in $\mathcal{A}$.
\end{proposition}

\begin{proof}
	Let $\mu: A^* \to M$ be the syntactic morphism of $L(\mathcal{A})$. 
	Suppose $\mathcal{P}_{\DA}$ is present in $\mathcal{A}$. Then there exists $u,v,p,q \in A^*$ such that $u = h(A_x)$, $v = h(x)$, $\alp(u) \subseteq \alp(v)$ and, without loss of generality, $pv^nuv^{n}q \in L(\mathcal{A})$ for all $n$ while $pv^{n}q \notin L(\mathcal{A})$ for all $n$. We get $\mu(v)^{\omega}\mu(u)\mu(v)^{\omega} \neq \mu(v)^{\omega}$, which by Lemma \ref{lem:DAproperty} implies $M \notin \DA$.

	On the other hand, suppose $M \notin \DA$. Then there are $u,v,p,q \in A^*$ such that, without loss of generality, $p(vu)^{\omega_Mn_1}u(vu)^{\omega_Mn_2}q \in L$ while $p(vu)^{\omega_Mn_3}q \notin L$ for all $n_1,n_2,n_3$. Define $h(x) = (vu)^{2\omega_M\eta}$ and $h(A_x) = u(vu)^{\omega_M\eta}$. By the definition of $\eta$, it follows that choosing $g(j) = \iota \cdot p(vu)^{\omega\eta}$, $g(k) = g(j) \cdot h(A_x)$ is a well defined homomorphism of semi-automata. Since $g(j) \not \semeq{\mathcal{A}} g(k)$, we have the desired pattern.
\end{proof}

In Section \ref{sec:reverse} we give a formal treatment of patterns for reverse-DFAs (which is necessary in order to deal with Carton-Michel automata, see Section \ref{sec:PatternsForCartonMichelAutomata}). We note that the following results, which are given for DFAs, have symmetric versions for reverse-DFAs.

In general, the presence of patterns is a feature of the particular automata, and not the language. For example, consider the following two automata recognizing the same language:
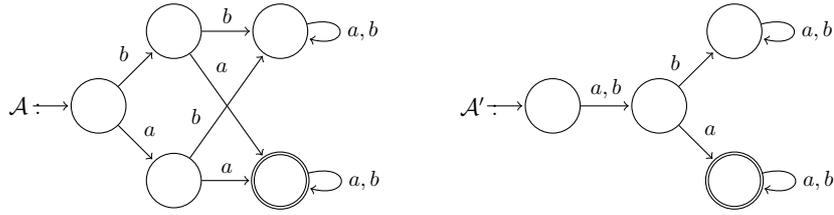
\begin{equation*}
	\begin{tikzpicture}[shorten >=1pt,node distance=1.4cm,on grid,auto,initial text = {},every node/.style={scale = 0.75}] 
		\node [left=of q_0] {\qquad\quad\large$\mathcal{A}:$};
		\node[state, initial] (q_0) {}; 
		\node[state] (q_1) [below right=of q_0] {}; 
		\node[state] (q_3) [above right=of q_0] {}; 
		\node[state,accepting] (q_2) [right=of q_1] {}; 
		\node[state] (q_4) [right=of q_3] {}; 
	    \path[->] (q_0) edge  node {$a$} (q_1);
	    \path[->] (q_0) edge  node {$b$} (q_3);
	    \path[->] (q_1) edge  node {$a$} (q_2);
	    \path[->] (q_3) edge  node {$b$} (q_4);
	    \path[->] (q_3) edge[near start]  node {$a$} (q_2);
	    \path[->] (q_1) edge[near start]  node {$b$} (q_4);
	    \path[->] (q_2) edge[loop right]  node {$a,b$} (q_2);
	    \path[->] (q_4) edge[loop right]  node {$a,b$} (q_4);
	\end{tikzpicture}
	\begin{tikzpicture}[shorten >=1pt,node distance=1.4cm,on grid,auto,initial text = {},every node/.style={scale = 0.75}] 
		\node[state, initial] (q_0) {}; 
		\node[state] (q_1) [right=of q_0] {}; 
		\node[left=of q_0] {\quad\qquad\large$\mathcal{A}':$};
		\node[state,accepting] (q_2) [below right=of q_1] {}; 
		\node[state] (q_4) [above right=of q_1] {}; 
	    \path[->] (q_0) edge  node {$a,b$} (q_1);
	    \path[->] (q_1) edge  node {$a$} (q_2);
	    \path[->] (q_1) edge[near start]  node {$b$} (q_4);
	    \path[->] (q_2) edge[loop right]  node {$a,b$} (q_2);
	    \path[->] (q_4) edge[loop right]  node {$a,b$} (q_4);
	\end{tikzpicture}
\end{equation*}
Let $\mathcal{P} = \left( \mathcal{S},j \neq k \right)$ be given by:
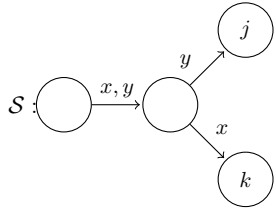
\begin{equation*}
	\begin{tikzpicture}[shorten >=1pt,node distance=1.4cm,on grid,auto,initial text = {},every node/.style={scale = 0.75}] 
		\node [left=of q_0] {\qquad\qquad\qquad\large$\mathcal{S}:$};
		\node[state] (q_0) {}; 
		\node[state] (q_1) [right=of q_0] {}; 
		\node[state] (q_2) [below right=of q_1] {$k$}; 
		\node[state] (q_4) [above right=of q_1] {$j$}; 
	    \path[->] (q_0) edge  node {$x,y$} (q_1);
	    \path[->] (q_1) edge  node {$x$} (q_2);
	    \path[->] (q_1) edge[near start]  node {$y$} (q_4);
	\end{tikzpicture}
\end{equation*}
We note that $\mathcal{P}$ is present in $\mathcal{A}'$ but not in $\mathcal{A}$ (see also \cite[Example 3.4]{KlimaPolak20jalc}). 
We are interested in patterns which are indeed a feature of the language rather than the particular automata, and thus we make the following definition. It is essentially the same as the $\mathsf{H}$-invariant configurations of Kl\'ima and Pol\'ak~\cite{KlimaPolak20jalc}.

\begin{definition}
	A (subword-)pattern is a \emph{language pattern} if for all $\mathcal{A},\mathcal{A}'$ such that $L(\mathcal{A}) = L(\mathcal{A}')$, we have $\mathcal{P}$ present in $\mathcal{A}$ if and only if it is present in $\mathcal{A}'$.
\end{definition}

\begin{definition}
	Let $\mathbf{P}$ be a collection of language patterns. Then $\genby{\mathbf{P}}$ is the set of languages $L(\mathcal{A})$ such that $\mathcal{A}$ does not have any of the patterns $\mathcal{P} \in \mathbf{P}$. For a finite set of patterns $\{\mathcal{P}_1,\dots,\mathcal{P}_n\}$, we use the notation $\genby{\mathcal{P}_1,\dots,\mathcal{P}_n}$ rather than $\genby{\{\mathcal{P}_1,\dots,\mathcal{P}_n\}}$.
\end{definition}

We show that language patterns gives rise to language varieties. This result, and the proof thereof, is analogous to that by Kl\'ima and Pol\'ak for $\mathsf{H}$-invariant configurations \cite{KlimaPolak20jalc}.

\begin{proposition}\label{prp:LanguagePatternsMakesLanguageVariety}
	Let $\mathbf{P}$ be a collection of language patterns. Then $\genby{\mathbf{P}}$ is a language variety.
\end{proposition}

\begin{proof}
	Since the class of varieties is closed under intersection, it is enough to show the statement for a single pattern $\mathcal{P} = \left( \mathcal{S}, j \not \leq k \right)$.

	Let $\mathcal{A} = \left( Q,A,\cdot,i,F \right)$ be an automaton accepting $L \in \genby{\mathcal{P}}$.
	We first consider the left-residual $u^{-1} L$. By setting $i' = i \cdot u$ we get $\mathcal{A} = \left( Q,A,\cdot,i',F \right)$ recognising $u^{-1}L$. It is clear that any pattern in $\mathcal{A}'$ is also in $\mathcal{A}$, so $u^{-1}L \in \genby{\mathcal{P}}$.

	For the right residual $Lv^{-1}$, let $F'$ be the set of states $j$ in $Q$ such that $j \cdot v \in F$. Let $\mathcal{A}' = \left( Q,A,\cdot,i,F' \right)$, then $\mathcal{A}'$ accepts $Lv^{-1}$. If $\mathcal{P}$ is present in $\mathcal{A}'$, then there exists witnesses $g,h$ and $w$ such that $g(j) \cdot w \in F'$ while $g(k) \cdot w \notin F'$. But then $g(j) \cdot wv \in F$ while $g(k) \cdot wv \notin F$. Hence $g,h$ witnesses that $\mathcal{P}$ is present in $\mathcal{A}$.

	Next, let $\mathcal{A}_1,\mathcal{A}_2$ be automata recognising $L_1,L_2 \in \genby{\mathcal{P}}$, and let $\mathcal{A}$ be the product automata with $F = \left\{ (\ell,m) \mid \ell \in F_1 \text{ and } m \in F_2 \right\}$. The homomorphisms $g : \mathcal{S} \to \mathcal{A}^h$ are given exactly by the pairs $g(n) = \left( g_1(n),g_2(n) \right)$ where $g_1: \mathcal{S} \to \mathcal{A}_1^h$, $g_2: \mathcal{S} \to \mathcal{A}_2^h$ and $h$ are arbitrary homomorphisms. Suppose that $\left( g_1(n),g_2(n) \right) \cdot w \in F$ while $\left( g_1(n'),g_2(n') \right) \cdot w \notin F$. Then without loss of generality $g_1(n) \cdot w \in F_1$, while $g_1(n') \cdot w \notin F_1$ showing that the pattern exists also in $\mathcal{A}_1$. A similar argument is applicable for the case when $F = \left\{ (\ell,m) \mid \ell \in F_1 \text{ or } m \in F_2 \right\}$.

	Finally, suppose $\mathcal{A} = (Q,A,\cdot,i,F)$ and let $f: B^* \to A^*$ be a homomorphism. The automata $\mathcal{A}^f = (Q,A,\cdot^f,i,F)$ recognises $f^{-1}(L(\mathcal{A}))$. Suppose it has the pattern $\mathcal{P}$, witnessed by a homomorphism $h: X^* \to B^*$ and a semiautomata homomorphism $g: \mathcal{S} \to \mathcal{A}^{f \circ h}$.
	Since $g(j) \not\semleq{\mathcal{A}^{f}} g(k)$, there is $u \in B^*$ such that $g(j) \cdot^f u \in F$, while $g(k) \cdot^f u \notin F$. It follows that $g(j) \cdot f(u) \in F$ while $g(k) \cdot f(u) \notin F$, and thus $f \circ h$ and $g$ are witnesses for the pattern being in $\mathcal{A}$. 
\end{proof}

We extend simple and balanced patterns to subword-patterns. If $\preceq$ is the identity, then 
conditions 
\itref{ddd:PrunableDefinition} and \itref{eee:PrunableDefinition} are trivial, and the definition reduces to that in~\cite{KlimaPolak20jalc}.\footnote{The reduction is up to a slight technical difference; we assume that every loop is preceded by an edge from a different state, which is not assumed in Kl\'ima and Pol\'ak. Any simple and balanced pattern in the sense of Kl\'ima and Pol\'ak yields an equivalent simple and balanced pattern in our sense by (if necessary) adding a transition (with a new variable) in front of the root.}

\begin{definition}\label{def:PrunableDefinition}
	Let $\mathcal{S} = (V,X,\circ)$ be a partial semiautomaton, and $\mathcal{P} = (\mathcal{S}, j \leq k)$ a subword-pattern. Then $\mathcal{P}$ is \emph{simple} if it is a tree after removing all self-loops.

	Let $\mathcal{L} = \left\{ x \in X \mid \ell \circ x = \ell \text{ for some $\ell \in V$} \right\}$ and let
	\begin{equation*}
		\mathcal{K}  = \left\{ (x,y) \in X \times \mathcal{L} \mid \ell \neq \ell \circ x = \ell \circ xy \text{ for some $\ell \in V$} \right\}.
	\end{equation*}
	I.e., $\mathcal{L}$ is the collection of variables which occurs as some loop in $\mathcal{S}$, and $\mathcal{K}$ is the collection of all pairs $(x,y)$ occurring together with $\ell \neq \ell'$ as follows:
	
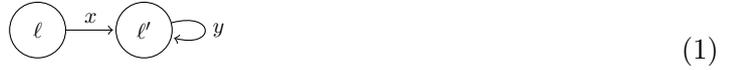
\begin{equation}\label{eqn:BalancedOccurrence}
	\begin{tikzpicture}[shorten >=1pt,node distance=1.4cm,on grid,auto,initial text = {},every node/.style={scale = 0.75}] 
		\node[state] (q_m1) {$\ell'$}; 
		\node[state] (q_m2) [left=of q_m1] {$\ell$}; 
	    \path[->] (q_m1) edge[loop right]  node {$y$} (q_m1);
	    \path[->] (q_m2) edge  node {$x$} (q_m1);
	\end{tikzpicture}
	\end{equation}
	The pattern $\mathcal{P}$ is \emph{balanced} if
	\begin{enumerate}[\itshape(i)]
		\item for all $y \in \mathcal{L}$, there exists $x \in X$ such that $(x,y) \in \mathcal{K}$, \label{aaa:PrunableDefinition}
		\item for all $(x,y) \in \mathcal{K}$, if $(x,y') \in \mathcal{K}$ then $y' = y$, \label{bbb:PrunableDefinition}
		\item for all $(x,y) \in \mathcal{K}$, if $\ell \circ x$ is defined, then $\ell \circ x y = \ell \circ x$. In other words, whenever $x$ occurs in $\mathcal{S}$, then it occurrs together with $y$ as in (\ref{eqn:BalancedOccurrence}), \label{ccc:PrunableDefinition}
		\item for all $y \in \mathcal{L}$, if $y \preceq z$, then $z \in \mathcal{L}$, \label{ddd:PrunableDefinition}
		\item for all $(x,y) \in \mathcal{K}$, if $x \preceq z$ for $x \neq z$ then $y \preceq z$. \label{eee:PrunableDefinition}
	\end{enumerate}
\end{definition}

The rest of the section is devoted to generalising a result of Kl\'ima and Pol\'ak that simple and balanced patterns are language patterns.\footnote{Or that they are $\mathsf{H}$-invariant in the language of~\cite{KlimaPolak20jalc}} The generalisation is straight-forward and follows the same line of argument as that of~\cite{KlimaPolak20jalc}. 

\begin{definition}
	Let $\mathcal{P}$ be a balanced pattern on $X$. For a homomorphism $h: X^* \to A^*$, we define $h_n$ by:
	\begin{equation*}
		h_n(x) = 
		\begin{cases}
			(h(x))^{2n} & \text{ if }x \in \mathcal{L},
			\\ h(x)(h(y))^{n} & \text{ if } (x,y) \in \mathcal{K} \text{ and } x \neq y
			\\ h(x) & \text{ otherwise. }
		\end{cases}
	\end{equation*}
\end{definition}

This is well defined because of condition \itref{bbb:PrunableDefinition} in Definition \ref{def:PrunableDefinition}. 
By condition \itref{ddd:PrunableDefinition} and \itref{eee:PrunableDefinition}, we also have that if $h$ satisfies the subword property, i.e.\ $x \preceq y$ implies $h(x)$ is a subword of $h(y)$, then $h_n$ also satisfies this property.

The following lemma shows that for Simple and Balanced patterns, we have a lot of candidates for the witnesses $g$ and $h$ showing presence in the DFA $\mathcal{A}$. In particular, every homomorphism $h$ and every state $\ell$ in $\mathcal{A}$ gives rise to such a candidate; all that is left is to check whether $g(j) \not \semleq{\mathcal{A}} g(k)$ (respectively $g(j) \not \semeq{\mathcal{A}} g(k)$) for any such candidate.

\begin{lemma}\label{lem:PrunablePatternIsEverywhere}
	Let $\mathcal{A} = \left( Q,A,\cdot,i,F \right)$ be an automaton and let $\mathcal{P}$ be a simple and balanced pattern with underlying semiautomaton $\mathcal{S} = \left( V,X,\circ \right)$ and root $r \in V$. Let $h: X^* \to A^*$ and $\ell \in Q$ be arbitrary. Setting $g(r) = \ell$, $g(r \circ y) = \ell \cdot h_{\eta}(y)$ gives a well defined homomorphism of semiautomata $g: \mathcal{S} \to \mathcal{A}^{h_{\eta}}$.
\end{lemma}

\begin{proof}
	Let $\mathcal{S}' = \left( V,X,\circ' \right)$ be $\mathcal{S}$ with all loops removed. Since $\mathcal{S}'$ is a tree, there is for each state $\ell \in V$ a unique $x \in X^*$ such that $r \circ' x = \ell$. Thus, $g(\ell) = g(r \circ' x) = \ell \cdot h_{\eta}(x)$ is well defined for all $\ell$. All that is left is to show that every loop in $\mathcal{S}$ maps to a loop in $\mathcal{A}$. Let $y \in \mathcal{L}$, and $\ell \in V$ such that $\ell \cdot y = \ell$. By condition \itref{aaa:PrunableDefinition} in Definition \ref{def:PrunableDefinition}, we have $\ell' \in V$, $x \in X$ such that $\ell' \neq \ell$ and $\ell' \cdot x = \ell$. We have $g(\ell) = g(\ell') \cdot h(x)(h(y))^{\eta}$. By the definition of $\eta$, this implies that $(h(y))^{\eta}$, and thus $(h(y))^{2\eta}$ is a loop at $g(\ell)$.
\end{proof}

A common use case for the chain of function $h_n$ will be as follows. We have an automaton $\mathcal{A}$ with witnesses $g$ and $h$ showing the existence of some simple and balanced pattern. Now, we want to find a candidate witness in some other automaton $\mathcal{B}$, and show that it is indeed a witness. To facilitate such arguments, we rely on the fact that $h$ and $h_{\eta_{\mathcal{B}}}$ are both witnesses in $\mathcal{A}$.

\begin{lemma}\label{lem:WeCanChangeHToHn}
	Suppose $\mathcal{P} = \left( \mathcal{S},j \not \leq k \right)$ is a simple and balanced pattern and suppose $h$ and $g$ are such that $g: \mathcal{S} \to \mathcal{A}^h$ is a homomorphism. Then $g_n: \mathcal{S} \to \mathcal{A}^{h_n}$ defined by $g_n(\ell) = g(\ell)$ is also a homomorphism. In particular, if $g$ and $h$ are witnesses for $\mathcal{P}$ being present in $\mathcal{A}$, then $g_n$,$h_n$ are also witnesses for all $n$.
\end{lemma}

\begin{proof}
	We show that $g(\ell) \cdot h(x) = g(\ell) \cdot h_n(x)$ for all $\ell \in V$, $x \in X$, which implies the desired result. For $x$ such that neither $(x,y) \in \mathcal{K}$ nor $x \in \mathcal{L}$, there is nothing to show. Suppose $x$ is such that $(x,y) \in \mathcal{K}$, and let $\ell' = \ell \circ x$. By condition \itref{ccc:PrunableDefinition}, we have that $y$ is a loop at $\ell'$ and it follows that $h(y)$ needs to be a cycle at $g(\ell')$. Thus 
	\begin{equation*}
	g(\ell) \cdot h_n(x) = g(\ell) \cdot h(x)\left( h(y) \right)^{n} = g(\ell') \cdot \left( h(y) \right)^{n} = g(\ell') = g(\ell) \cdot h(x).
	\end{equation*}
	The argument for $x \in \mathcal{L}$ is similar.
\end{proof}

It follows from these two lemmas that every simple and balanced pattern is a language pattern.

\begin{proposition}[See {\cite[Proposition 3.8]{KlimaPolak20jalc}}]
	Every simple and balanced subword-pattern is a language pattern.
\end{proposition}

\begin{proof}
	Let $\mathcal{P} = \left( \mathcal{S}, j \not \leq k \right)$ be a simple and balanced pattern with $\mathcal{S} = \left( V,X,\circ \right)$. Let $\mathcal{A} = \left( Q,A,\cdot,i,F \right)$ and $\mathcal{A}' = \left( Q',A,\cdot',i',F' \right)$ both accept $L \subseteq A^*$. 
	Assume that $\mathcal{P}$ is present in $\mathcal{A}$, witnessed by $h: X \to A^*$ and $g: \mathcal{S} \to \mathcal{A}^h$. Let $r$ be the root of $\mathcal{P}$, and choose $p$ such that $g(r) = i \cdot p$. There exists $x,y \in X^*$ such that $r \circ x = j$, $r \circ y = k$. Furthermore, there exists $q$ such that $g(j) \cdot q \in F$ while $g(k) \cdot q \notin F$. Thus $ph(x)q \in L$ while $ph(y)q \notin L$. It follows by Lemma \ref{lem:WeCanChangeHToHn} that $ph_n(x)q \in L$ while $ph_n(y)q \notin L$ for all $n$. 

	Let $\eta = \eta_{\mathcal{A}'}$. By Lemma \ref{lem:PrunablePatternIsEverywhere}, we can find a homomorphism $g': \mathcal{S} \to \mathcal{A}'$ such that $g'(r) = i' \cdot' p$, and $g(r \circ z) = g(r) \cdot' h_{\eta}(z)$ for all $z \in X^*$. In particular, $g(j) = i' \cdot ph_{\eta}(x)$, $g(k) = i' \cdot ph_{\eta}(y)$. Since $i' \cdot ph_{\eta}(x)q \in F'$ and $i' \cdot ph_{\eta}(y)q \notin F'$, we have $g(j) \not \semleq{\mathcal{A}} g(k)$, showing existence of $\mathcal{P}$ in $\mathcal{A}'$. The result for type 1 patterns is analogous.
\end{proof}

\section{Hierarchies of Subword-Patterns}
\label{sec:Hierarchies}

In this section, we show how to use patterns which characterize a variety $\mathbf{V} \subseteq \DA$ to create new patterns characterizing $\mathbf{K} \malcev \mathbf{V}$, $\mathbf{D} \malcev \mathbf{V}$ and $\genmalcev{\mathbf{V}}$. Patterns characterizing $\mathbf{R}_m$, $\mathbf{L}_m$ and $\Si_m$ becomes an immediate corollary. We also give these patterns explicitly.

Given a pattern $\mathcal{P}$, we construct patterns $\mathcal{P}_k$, $\mathcal{P}_d$ and $\mathcal{P}_{kd}$. These are obtained by appending new states either at the root of $\mathcal{P}$ as in (\ref{eqn:AppendAtRoot}) below (for $\mathcal{P}_k$), at the two states which were compared in $\mathcal{P}$ as in (\ref{eqn:AppendAtEnd}) below (for $\mathcal{P}_d$), or both (for $\mathcal{P}_{kd}$).\vspace{1em}

\hspace{-2.4em}
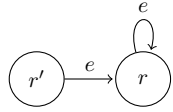
\begin{minipage}[c]{0.5\textwidth}
	\begin{equation}\label{eqn:AppendAtRoot}
	\begin{tikzpicture}[shorten >=1pt,node distance=1.4cm,on grid,auto,initial text = {},every node/.style={scale = 0.75}] 
		\node[state] (q_m1){$r$}; 
		\node[state] (q_m2) [left=of q_m1] {$r'$}; 
	    \path[->] (q_m1) edge[loop above]  node {$e$} (q_m1);
	    \path[->] (q_m2) edge  node {$e$} (q_m1);
	\end{tikzpicture}
  \end{equation}
\end{minipage}
\begin{minipage}[c]{0.5\textwidth}
	
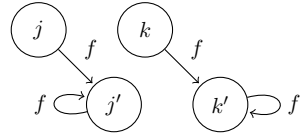
\begin{equation}\label{eqn:AppendAtEnd}
	\begin{tikzpicture}[shorten >=1pt,node distance=1.4cm,on grid,auto,initial text = {},every node/.style={scale = 0.75}] 
		\node[state] (q_0) {$j$}; 
		\node[state] (q_0p) [below right=of q_0] {$j'$}; 
		\node[state] (q_2) [right=of q_0] {$k$}; 
		\node[state] (q_2p) [below right=of q_2] {$k'$}; 
	    \path[->] (q_0) edge  node {$f$} (q_0p);
	    \path[->] (q_0p) edge[loop left]  node {$f$} (q_0p);
	    \path[->] (q_2) edge  node {$f$} (q_2p);
	    \path[->] (q_2p) edge[loop right]  node {$f$} (q_2p);
	\end{tikzpicture}
  \end{equation}
\end{minipage}
\vspace{1em}

When appending states as in (\ref{eqn:AppendAtEnd}), we compare $j'$ and $k'$ in the new pattern. The variables $e$ and $f$ are new, and defined to satisfy $x \preceq e,f$ for all variables $x$ of the original pattern $\mathcal{P}$. Formally, we have the following definition.

\begin{definition}
	Let $\mathcal{P} = (\mathcal{S},j \neq k)$ be a rooted pattern where $\mathcal{S} = (V,X,\circ)$ with the root $r$. Let $X_k = X \cup \left\{ e \right\}$ where $x \prec e$ for all $x \in X$, and let $V_k = V \cup \left\{ r' \right\}$. Let $\mathcal{S}_k = (V',X',\circ_k)$ where $r' \circ_k e = r$, $r \circ_k e = r$ and $\ell \circ_k x = \ell \circ x$ for all $\ell \in V$, $x \in X$ for which $\ell \circ x$ is defined. We define $\mathcal{P}_k = (\mathcal{S}_k,j \neq k)$.

	Next, let $X_d = X \cup \left\{ f \right\}$, $V_d = V \cup \left\{ j',k' \right\}$ and let $j \circ_d f = j'$, $k \circ_d f = k'$, $j' \circ_d f = j'$, $k' \circ_d f = k'$ and $\ell \circ_d x = \ell \circ x$ for all $\ell \in V$, $x \in X$ for which $\ell \circ x$ is defined. Then $\mathcal{S}_{d} = (V'',X',\circ_d)$ and $\mathcal{P}_d = (\mathcal{S}_d,j' \neq k')$.

	Finally, let $X_{kd} = X \cup \left\{ e,f \right\}$ where $x \prec e$, $x \prec f$ for all $x \in X$, and let $V_{kd} = V \cup \left\{ r',i',j' \right\}$. We define $r' \circ_{kd} e = r$, $r \circ_{kd} e = r$, $j \circ_{kd} f = j'$, $k \circ_{kd} f = k'$, $j' \circ_{kd} f = j'$, $k' \circ_{kd} f = k'$ and $\ell \circ_{kd} x = \ell \circ x$ for all $\ell \in V$, $x \in X$ for which $\ell \circ x$ is defined. Then $\mathcal{S}_{kd} = (V',X'',\circ_{kd})$ and $\mathcal{P}_{kd} = \left( \mathcal{S}_{kd},j' \neq k' \right)$.

	We make analogous definitions for type 2 patterns $\mathcal{P} = (\mathcal{S}, j \not \leq k)$. 
\end{definition}

As an example, we consider the simple and balanced pattern $\mathcal{P}_{\DA}'$ obtained by adding a root and a transition $y$ going into the state $j$. Let $y$ and $A_y$ as in $\mathcal{P}_{\DA}'$. The pattern $(\mathcal{P}_{\DA}')_{kd}$ is given by
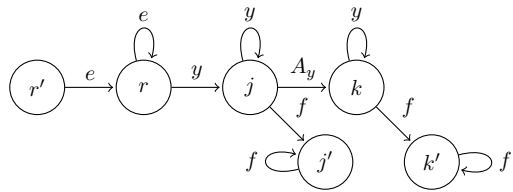
\begin{equation*}
	\begin{tikzpicture}[shorten >=1pt,node distance=1.4cm,on grid,auto,initial text = {},every node/.style={scale = 0.75}] 
		\node[state] (q_0) {$j$}; 
		\node[state] (q_m1) [left=of q_0] {$r$}; 
		\node[state] (q_m2) [left=of q_m1] {$r'$}; 
		\node[state] (q_0p) [below right=of q_0] {$j'$}; 
		\node[state] (q_2) [right=of q_0] {$k$}; 
		\node[state] (q_2p) [below right=of q_2] {$k'$}; 
	    \path[->] (q_m1) edge[loop above]  node {$e$} (q_m1);
	    \path[->] (q_0) edge[loop above]  node {$y$} (q_0);
	    \path[->] (q_0) edge  node {$A_y$} (q_2);
	    \path[->] (q_m1) edge  node {$y$} (q_0);
	    \path[->] (q_m2) edge  node {$e$} (q_m1);
	    \path[->] (q_2) edge[loop above]  node {$y$} (q_2);
	    \path[->] (q_0) edge  node {$f$} (q_0p);
	    \path[->] (q_0p) edge[loop left]  node {$f$} (q_0p);
	    \path[->] (q_2) edge  node {$f$} (q_2p);
	    \path[->] (q_2p) edge[loop right]  node {$f$} (q_2p);
	\end{tikzpicture}
\end{equation*}
It is straightforward to show that $(\mathcal{P}_{\DA})_{kd}$ is in fact equivalent to $\mathcal{P}_{\DA}$.

Note that if $\mathcal{P}$ is simple and balanced, then the patterns $\mathcal{P}_k$, $\mathcal{P}_d$ and $\mathcal{P}_{kd}$ are all simple and balanced. The constructions also preserve another property. We want to consider patterns where the alphabet of one path is a subset of the other (for type 2 patterns), or where they are the same (for type 1 patterns). This is ensured by the following property.

\begin{definition}
	Let $\mathcal{P} = \left( \mathcal{S},j \not \leq k \right)$ be a simple pattern such that whenever $x$ is on the path from $r$ to $j$, then there exists $y$ on the path from $r$ to $k$ such that $x \preceq y$. We say that  $\mathcal{P}$ is \emph{one-alphabeted}. 
	If $\mathcal{P} = \left( \mathcal{S}, j \neq k \right)$, then it is one-alphabeted if both the above holds and the for all $x$ on a path from $r$ to $k$, there is $y$ on the path from $r$ to $j$ such that $x \preceq y$.
\end{definition}

We show that if there is a collections of simple, balanced and one-alphabeted patterns characterizing monoid varieties inside of $\DA$, then these constructions can be used to obtain pattern characterizations for Malcev products with $\mathbf{K}$ and $\mathbf{D}$ and varieties constructed using the $\preceq_{\mathbf{KD}}$-relation. This requires the following two lemmas.

\begin{lemma}\label{lem:InductivePatternsAndMalcevForLanguages}
	Let $\mathcal{A} = \left( Q,A,\cdot,i,F \right)$ be a DFA and let $L = L(\mathcal{A})$ have the syntactic morphism $\mu: A^* \to M \in \DA$.
	Let $\mathcal{P}$ be a simple and balanced pattern. Then the following holds:
	\begin{enumerate}[(i)]
		\item If $\mathcal{P}_k$ is present in $\mathcal{A}$, then $\mathcal{P}$ is present in the Cayley-graph of $\faktor{M}{\sim_{\mathbf{K}}}$,
		\item If $\mathcal{P}_d$ is present in $\mathcal{A}$, then $\mathcal{P}$ is present in the Cayley-graph of $\faktor{M}{\sim_{\mathbf{D}}}$,
		\item If $\mathcal{P}_{kd}$ is present in $\mathcal{A}$, then $\mathcal{P}$ is present in the Cayley-graph of $\faktor{M}{\preceq_{\mathbf{KD}}}$,
	\end{enumerate}
	where we define presence in the Cayley-graph to mean that it is possible to make a choice of final states $F$ such that the pattern is present in the corresponding automata.
\end{lemma}

\begin{proof}
	Let $\mathcal{C} = \left( \faktor{M}{\sim_{\mathbf{K}}},A,\circ \right)$ be the Cayley-graph of $\faktor{M}{\sim_{\mathbf{K}}}$. Suppose $\mathcal{P}_k$ is present in $\mathcal{A}$ witnessed by $h: X'^* \to A^*$ and $g: \mathcal{S}_k \to \mathcal{A}^h$. By Lemma \ref{lem:WeCanChangeHToHn}, $h_{\eta_{\mathcal{C}}}: X'^* \to A^*$ together with $g$ is also a witness.
	Let $r \in V$ be the root of $\mathcal{P}$, $r' \in V'$ be the root of $\mathcal{P}_k$ and let $x,y \in X'^*$ such that $r \cdot' x = j$, $r \cdot' y = k$ where $\cdot'$ is the transition function of the underlying semiautomaton of $\mathcal{P}$. Choose $p$ such that $g(r') = i \cdot p$. Then there exist $q$ such that,without loss of generality $ph_{\eta_{\mathcal{C}}}(e)^{\omega_M}h_{\eta_{\mathcal{C}}}(x)q \in L$ while $ph_{\eta_{\mathcal{C}}}(e)^{\omega_M}h_{\eta_{\mathcal{C}}}(y)q \notin L$. Since $M \in \DA$ and $\alp(h_{\eta_{\mathcal{C}}}(x)) \subseteq \alp(h_{\eta_{\mathcal{C}}}(e))$, Lemma \ref{lem:DAproperty} gives $\mu(h_{\eta_{\mathcal{C}}}(e^{\omega_M}x)) \Jeq \mu(h_{\eta_{\mathcal{C}}}(e^{\omega_M}))$. Thus we have $\mu(h_{\eta_{\mathcal{C}}}(x)) \not\sim_{\mathbf{K}} \mu(h_{\eta_{\mathcal{C}}}(y))$.

	The latter implies that $1 \circ h_{\eta_{\mathcal{C}}}(x) \neq 1 \circ h_{\eta_{\mathcal{C}}}(y)$ in the Cayley-graph of $\faktor{M}{\sim_{\mathbf{K}}}$ where $1$ is the unit of $\faktor{M}{\sim_{\mathbf{K}}}$. Since $\mathcal{P}$ is simple and balanced, we can use Lemma \ref{lem:PrunablePatternIsEverywhere} to extend $g'(r) = 1$ to a homomorphism $g': \mathcal{S} \to \mathcal{C}^{h_{\eta_{\mathcal{C}}}}$. Setting $\mathcal{A}' = \left(\faktor{M}{\sim_{\mathbf{K}}},A,\circ,1,\left\{ 1 \circ h_{\eta_{\mathcal{C}}}(x) \right\} \right)$ gives $g'(i) \not \semeq{\mathcal{A}'} g'(j)$ which shows that $\mathcal{P}$ is present in $\mathcal{C}$. The other cases are similar. 
\end{proof}

\begin{lemma}\label{lem:InductivePatternsAndMalcevForMonoids}
	Let $\mathcal{A}$ be a DFA, and suppose $\mu: A^* \to M$ is its syntactic morphism. Suppose that $\mathcal{P}$ is a simple, balanced and one-alphabeted pattern. If $\mathcal{P}$ is type 1, then
	\begin{enumerate}[(i)]
		\item If $\mathcal{P}_k$ is not present in $\mathcal{A}$, then $\mathcal{P}$ is not present in any automata $\mathcal{B}$ accepting any $L$ recognised by $\faktor{M}{\sim_{\mathbf{K}}}$,\label{aaa:InductivePatternsAndMalcevForMonoids}
		\item If $\mathcal{P}_d$ is not present in $\mathcal{A}$, then $\mathcal{P}$ is not present in any automata $\mathcal{B}$ accepting any $L$ recognised by $\faktor{M}{\sim_{\mathbf{D}}}$,\label{bbb:InductivePatternsAndMalcevForMonoids}
	\setcounter{continueListCounter}{\value{enumi}}
	\end{enumerate}
	and if $\mathcal{P}$ is either type 1 or type 2, then
	\begin{enumerate}[(i)]
	\setcounter{enumi}{\value{continueListCounter}}
		\item If $\mathcal{P}_{kd}$ is not present in $\mathcal{A}$, then $\mathcal{P}$ is not present in any automata $\mathcal{B}$ accepting any $L$ recognised by $\faktor{M}{\preceq_{\mathbf{KD}}}$.\label{ccc:InductivePatternsAndMalcevForMonoids}
	\end{enumerate}
\end{lemma}

\begin{proof}
	We show the result \itref{aaa:InductivePatternsAndMalcevForMonoids}, with \itref{bbb:InductivePatternsAndMalcevForMonoids} and \itref{ccc:InductivePatternsAndMalcevForMonoids} being similar. Let $\mathcal{P} = \left( \mathcal{S}, j \neq k \right)$, $r$ be the root of $\mathcal{P}$, and $r'$ the root of $\mathcal{P}_k$.
	Suppose $\mathcal{P}$ is present in some $\mathcal{B}$ with $h,g$ as witnesses. By Lemma \ref{lem:WeCanChangeHToHn}, we can also use $h_{\eta_{\mathcal{A}}}$ as a witness.
	Let $x,y \in X^*$ be such that $j = r \circ x$ and $k = r \circ y$. Let $u = h_{\eta_{\mathcal{A}}}(x)$, $v = h_{\eta_{\mathcal{A}}}(y)$.
	Since $\mathcal{P}$ is one-alphabeted, it follows that $\alp(u) = \alp(v)$.

	Since $g(r) \cdot u = g(j) \not \semeq{\mathcal{B}} g(k) = g(r) \cdot v$, it follows from the minimality of the syntactic morphism that $\mu(u) \not \sim_{\mathbf{K}} \mu(v)$.
	This implies that there exists $w$ such that either $v$ or $u$ is a factor of it and such that for all $n$, we have $pw^{n\omega_M}uq \in L(\mathcal{A}) \Leftrightarrow pw^{n\omega_M}vq \notin L(\mathcal{A})$ for some $p$ and $q$. In particular, since $u$ and $v$ have the same alphabet, we can choose $w$ such that both $u$ and $v$ are subwords of it.
	
	Let $h'(e) = w^{\omega_M}$, $h'(z) = h(z)$ for all $z \in X$. Using Lemma \ref{lem:PrunablePatternIsEverywhere}, we set $g'(r') = i \cdot p$ and get a well defined homomorphism $g': \mathcal{S}_k \to \mathcal{A}^{h_{\eta_{\mathcal{A}}}'}$. We get 
	\begin{equation*}
	g'(j) = g'(r') \cdot h_{\eta_{\mathcal{A}}}'(ex) = i \cdot pw^{\eta_{\mathcal{A}}\omega_M}u \not \semeq{\mathcal{A}} i \cdot pw^{\eta_{\mathcal{A}}\omega_M}v = g'(r') \cdot h_{\eta_{\mathcal{A}}}'(ey) = g'(k),
	\end{equation*}
	showing that the pattern $\mathcal{P}_k$ is present in $\mathcal{A}$. 
\end{proof}

Combining these lemmas yields the following theorem, which is the main result of this section. 

\begin{theorem}\label{thm:PatternsForMalcevProducts}
	Let $\mathbf{P}$ be a collection of simple, balanced and one-alphabeted patterns with $\mathcal{P}_{\DA}' \in \mathbf{P}$. Suppose $\mathcal{V} = \genby{\mathbf{P}}$. If all patterns in $\mathbf{P}$ are type 1, then
	\begin{enumerate}[(i)]
		\item the language variety corresponding to $\mathbf{K} \malcev \mathbf{V}$ is $\genby{\mathbf{P}_k}$,\label{aaa:PatternsForMalcevProducts}
		\item the language variety corresponding to $\mathbf{D} \malcev \mathbf{V}$ is $\genby{\mathbf{P}_d}$,\label{bbb:PatternsForMalcevProducts}
	\setcounter{continueListCounter}{\value{enumi}}
	\end{enumerate}
	and for $\mathbf{P}$ containing any combination of type 1 and type 2 patterns, we have
	\begin{enumerate}[(i)]
	\setcounter{enumi}{\value{continueListCounter}}
		\item the language variety corresponding to $\genmalcev{\mathbf{V}}$ is $\genby{\mathbf{P}_{kd}}$,\label{ccc:PatternsForMalcevProducts}
	\end{enumerate}
\end{theorem}

\begin{proof}
	We again show \itref{aaa:InductivePatternsAndMalcevForMonoids}, with \itref{bbb:InductivePatternsAndMalcevForMonoids} and \itref{ccc:InductivePatternsAndMalcevForMonoids} being analogous. 
	Consider $\mathcal{A}$ and let $M$ be the syntactic monoid of $L(\mathcal{A})$. 
	Suppose $\mathcal{A}$ has one of the patterns $\mathcal{P}_k \in \mathbf{P}_k$. Either it is $(\mathcal{P}_{\DA})_k$ which is equivalent to $\mathcal{P}_{\DA}$. Then $M \notin \DA$ and thus $M \notin \mathbf{K} \malcev \mathbf{V}$ since $\mathbf{K} \malcev \DA = \DA$.
	If it is not $\mathcal{P}_{\DA}$, then $M \in \DA$ and we can use Lemma \ref{lem:InductivePatternsAndMalcevForLanguages} to find a language $L'$ recognised by $\faktor{M}{\sim_{\mathbf{K}}}$ such that $L' \notin \genby{\mathcal{P}}$. It follows that $\faktor{M}{\sim_{\mathbf{K}}} \notin \mathbf{V}$, and thus $M \notin \mathbf{K} \malcev \mathbf{V}$.

	On the other hand, suppose that $\mathcal{A}$ has none of the patterns $\mathcal{P}_k \in \mathbf{P}_k$. Let $M$ be the syntactic monoid of $L(\mathcal{A})$. By Lemma \ref{lem:InductivePatternsAndMalcevForMonoids}, none of the languages recognised by $\faktor{M}{\sim_{\mathbf{K}}}$ has any of the patterns $\mathcal{P} \in \mathbf{P}$. For every such language $L_i$, let $M_i$ be the corresponding syntactic monoid. We have that $M_i \in \mathbf{V}$, and thus $\faktor{M}{\sim_{\mathbf{K}}} \in \mathbf{V}$. This shows that $M \in \mathbf{K} \malcev \mathbf{V}$.
\end{proof}

The explicit patterns for $\mathbf{R}_m$, $\mathbf{L}_m$ and $\mathbf{Si}_m$ all build on the same class of directed graphs. However, the orderings of the variables are different.

\begin{definition}
	For $m \geq 1$, we define the following sets of variables:
	\begin{itemize}
		\item $X_m = \left\{ x,e_{1},\dots,e_{\floor{m/2}},f_{1},\dots,f_{\floor{(m-1)/2}} \right\}$ with  $x \preceq e_i \preceq f_i \preceq e_{i+1}$,
		\item $Y_m = \left\{ x,e_1,\dots,e_{\floor{(m-1)/2}},f_1,\dots, f_{\floor{m/2}} \right\}$ with  $x \preceq f_i \preceq e_i \preceq f_{i+1}$,
		\item $Z_m = \left\{ x,e_1,\dots,e_{m-1},f_1,\dots,f_{m-1} \right\}$ with  $x \preceq y$ for all $y \in Z_m$ and $z_i \preceq z_{i+1}$ for $z_i \in \left\{ e_i,f_i \right\}$, $z_{i+1} \in \left\{ e_{i+1},f_{i+1} \right\}$.
	\end{itemize}
	Let $\mathcal{S}^X_m$ (resp. $\mathcal{S}^Y_m$, $\mathcal{S}^Z_m$) have the following structure, where $x,e_i,f_{i'} \in X_m$ (resp.\ in $Y_m$, $Z_m$) and $\ell$ and $\ell'$ are chosen to match the maximal $e_i$ and $f_{i'}$ respectively.

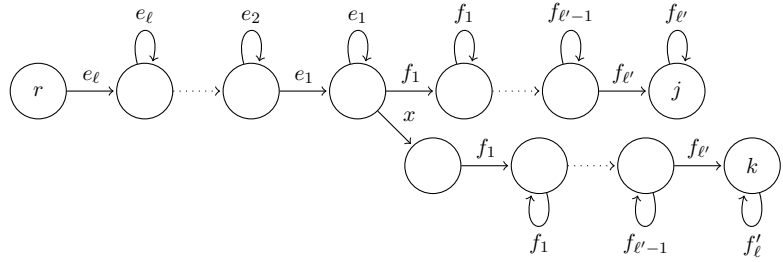
\begin{equation*}
	\begin{tikzpicture}[shorten >=1pt,node distance=1.4cm,on grid,auto,initial text = {},every node/.style={scale = 0.75}] 
	   \node[state] (q_0) {}; 
	\node[state] (q_m1) [left=of q_0] {}; 
	\node[state] (q_m2) [left=of q_m1] {}; 
	\node[state] (q_m3) [left=of q_m2] {$r$}; 
	   \node[state] (q_1) [below right=of q_0] {}; 
	   \node[state] (q_2) [right=of q_0] {}; 
	   \node[state] (q_3) [right=of q_1] {}; 
	   \node[state] (q_4) [right=of q_2] {}; 
	   \node[state] (q_5) [right=of q_3] {}; 
	   \node[state] (q_6) [right=of q_4] {$j$}; 
	   \node[state] (q_7) [right=of q_5] {$k$}; 
	   \path[->] (q_m3) edge  node {$e_{\ell}$} (q_m2);
	   \path[->] (q_m2) edge[dotted] node {} (q_m1);
	   \path[->] (q_m2) edge[loop above]  node {$e_{\ell}$} (q_m2);
	   \path[->] (q_m1) edge[loop above]  node {$e_2$} (q_m1);
	   \path[->] (q_m1) edge  node {$e_{1}$} (q_0);
	   \path[->] (q_0) edge[loop above]  node {$e_1$} (q_0);
	    \path[->] (q_0) edge  node {$x$} (q_1);
	    \path[->] (q_2) edge[loop above]  node {$f_1$} (q_2);
	    \path[->] (q_3) edge[loop below]  node {$f_1$} (q_3);
	    \path[->] (q_0) edge  node {$f_1$} (q_2);
	    \path[->] (q_1) edge  node {$f_1$} (q_3);
	    \path[->] (q_4) edge[loop above]  node {$f_{\ell'-1}$} (q_5);
	    \path[->] (q_5) edge[loop below]  node {$f_{\ell'-1}$} (q_4);
	    \path[->] (q_2) edge[dotted]  node {} (q_4);
	    \path[->] (q_3) edge[dotted]  node {} (q_5);
	    \path[->] (q_6) edge[loop above]  node {$f_{\ell'}$} (q_6);
	    \path[->] (q_7) edge[loop below]  node {$f_\ell'$} (q_7);
	    \path[->] (q_4) edge  node {$f_{\ell'}$} (q_6);
	    \path[->] (q_5) edge  node {$f_{\ell'}$} (q_7);
	\end{tikzpicture}
\end{equation*}

\noindent
Then
\begin{itemize}
	\setlength{\itemsep}{2pt}
	\item $\mathcal{P}^{\mathbf{R}}_m = \left( \mathcal{S}^X_m, j \neq k \right)$ for even $m \geq 2$, $\mathcal{P}^{\mathbf{R}}_m = \left( \mathcal{S}^Y_m, j \neq k \right)$ for odd $m \geq 3$,
	\item $\mathcal{P}^{\mathbf{L}}_m = \left( \mathcal{S}^Y_m, j \neq k \right)$ for even $m \geq 2$, $\mathcal{P}^{\mathbf{L}}_m = \left( \mathcal{S}^X_m, j \neq k \right)$ for odd $m \geq 3$,
	\item $\mathcal{P}^{\mathbf{Si}}_m = \left( \mathcal{S}^Z_m, j \not \leq k \right)$ for $m \geq 1$.
\end{itemize}
\end{definition}

Before showing that these patterns characterise the corresponding varieties, we show the following lemma, which allows us to use Theorem \ref{thm:PatternsForMalcevProducts} without adding $\mathcal{P}_{\DA}$ explicitly.

\begin{lemma}\label{lem:DAImpliesRLandSi}
	Let $\mathcal{A} = \left( Q,A,\cdot,i,F \right)$ be a DFA. Suppose $\mathcal{P}_{\DA}'$ is present in $\mathcal{A}$, then $\mathcal{P}^{\mathbf{R}}_m$, $\mathcal{P}^{\mathbf{L}}_m$ and $\mathcal{P}^{\Si}_m$ are present in $\mathcal{A}$ for all $m$.
\end{lemma}

\begin{proof}
	Proving the presence of $\mathcal{P}^\mathbf{R}_m$ and $\mathcal{P}^\mathbf{L}_m$ is straightforward. Since $\mathcal{P}_{\DA}'$ and $\mathcal{P}_{\DA}$ are equivalent, we may suppose $\mathcal{P}_{\DA}$ is present, with witnesses $h',g'$. To distinguish the variables, we prime the names of variables and states from $\mathcal{P}_{\DA}$. We define $h(e_i) = h(f_i) = h'(x')$ for all $i$, and $h(x) = h'(A_x')$. Defining $g(r) = g'(j')$ and extending it to a homomorphism of partial semi-DFAs gives $h$ and $g$ witnessing the presence of $\mathcal{P}^\mathbf{R}_m$ and $\mathcal{P}^\mathbf{L}_m$.

	If $g'(j') \not \leq g'(k')$ the same approach works for $\mathcal{P}^{\Si}_m$. However, if we only have $g'(k') \not \leq g'(j')$ we require some more work. Let $M$ be the syntactic monoid of $L(\mathcal{A})$. Since $M \notin \DA$, we have $p,q,u,v \in A^*$ such that $p(uv)^{\omega n_1}v(uv)^{\omega n_2}q \in L$ while $p(uv)^{\omega n_3}q \notin L$ for all $n_1,n_2,n_3$ or vice versa. In the latter case, we have the pattern $\mathcal{P}_{\DA}$ with $g'(j') \not \leq g'(k')$ so we consider the former case.

	We define $h(e_n) = (uv)^{\eta}$, $h(f_n) = (vu)^{\eta}$ and $h(x) = u(vu)^{\eta - 1}$ for all $n$. Setting $g(r) = i \cdot p$ gives a homomorphism $g: \mathcal{S}^Z_m \to \mathcal{A}$. All that is left is to show $g(j) \not \leq g(k)$. We note that $g(j) = i \cdot p(uv)^{\eta}(vu)^{\eta} = i \cdot p(uv)^{\omega\eta}v(uv)^{\omega\eta-1}u$ and $g(k) = i \cdot p(uv)^{\eta}u(vu)^{\eta-1}(vu)^{\eta} = i \cdot p(uv)^{\omega\eta-1}u$. We have $g(j) \cdot vq = i \cdot p(uv)^{\omega\eta}v(uv)^{\omega\eta-1}uvq = i \cdot p(uv)^{\omega\eta}v(uv)^{\omega\eta}q \in L$ and $g(k) \cdot vq = i \cdot p(uv)^{\omega\eta-1}uvq = i \cdot p(uv)^{\omega\eta}q \notin L$ giving the desired result.
\end{proof}

\begin{corollary}\label{cor:PatternsForLevels}
	Let $\mathcal{A}$ be a DFA, and let $M$ be the syntactic monoid of $L(\mathcal{A})$. Then the following holds:
	\begin{enumerate}[(i)]
	\setlength{\itemsep}{2pt}
	\item $M \in \mathbf{R}_m$ if and only if $L(\mathcal{A}) \in \genby{\mathcal{P}^{\mathbf{R}}_{m}}$,\label{aaa:PatternsForLevels}
	\item $M \in \mathbf{L}_m$ if and only if $L(\mathcal{A}) \in \genby{\mathcal{P}^{\mathbf{L}}_{m}}$,\label{bbb:PatternsForLevels}
	\item $M \in \Si_m$ if and only if $L(\mathcal{A}) \in \genby{\mathcal{P}^{\Si}_{m}}$.\label{ccc:PatternsForLevels}
	\end{enumerate}
\end{corollary}

\begin{proof}
	We proceed by induction on $m$. The inductive step uses Theorem \ref{thm:PatternsForMalcevProducts}, which require the presence of the pattern $\mathcal{P}_{\DA}'$. However, by Lemma \ref{lem:DAImpliesRLandSi}, it follows that $\genby{\mathcal{P}_{m}^{\mathbf{R}}} = \genby{\mathcal{P}_{m}^{\mathbf{R}}, \mathcal{P}_{\DA}'}$, $\genby{\mathcal{P}_{m}^{\mathbf{L}}} = \genby{\mathcal{P}_{m}^{\mathbf{L}}, \mathcal{P}_{\DA}'}$ and $\genby{\mathcal{P}_{m}^{\Si}} = \genby{\mathcal{P}_{m}^{\Si}, \mathcal{P}_{\DA}'}$. Thus we can without loss of generality assume the collections contain $\mathcal{P}_{\DA}'$ making Theorem \ref{thm:PatternsForMalcevProducts} applicable.
	We show the base case $\mathcal{R} = \genby{\mathcal{P}^{\mathbf{R}}_2} = \genby{\mathcal{P}^{\mathbf{R}}_2,\mathcal{P}_{\DA}'}$. The base case for $\mathcal{P}^{\mathbf{L}}_2$ is analogous and for $\mathcal{P}^{\Si}_1$ it is trivial. 

	Suppose $L(\mathcal{A}) \notin \genby{\mathcal{P}^{\mathbf{R}}_1}$. If $M \notin \DA$ then in particular $M \notin \mathbf{R}$, so we assume $M \in \DA$.
	Thus, we suppose that $\mathcal{P}^{\mathbf{R}}_1$ is present in $\mathcal{A}$ with $g$ and $h$ as the witnesses. Let $\omega$ be the idempotent power of $M$. Then there exists $p,q$ such that $ph(e_1)^{\omega}h(x)q \in L(\mathcal{A}) \Leftrightarrow ph(e_1)^{\omega}q \notin L(\mathcal{A})$ where $h(x)$ is a subword of $h(e_1)$. However, since $M \notin \DA$, we have $p'h(e_1)^{\omega}q' \in L(\mathcal{A})$ if and only if $p'h(e_1)^{\omega}h(x)h(e_1)^{\omega}q' \in L(\mathcal{A})$ for all $p',q' \in A^*$. Thus
  \begin{alignat*}{2}
	  ph(e_1)^{\omega}(h(x)h(e_1)^{\omega})^{\omega}h(x) q \in L(\mathcal{A}) 	  \quad \Leftrightarrow \quad 
	  & ph(e_1)^{\omega}h(x) q \in L(\mathcal{A}) 
	  \\  \Leftrightarrow \quad & ph(e_1)^{\omega}q \notin L(\mathcal{A})
	  \\ \Leftrightarrow \quad & ph(e_1)^{\omega}(h(x)h(e_1)^{\omega})^{\omega}q \notin L(\mathcal{A}),
  \end{alignat*}	
  showing that $M \notin \mathbf{R} = \mathbf{R}_2$.
  
  For the other direction, suppose $M \notin \mathbf{R}$. We then have words $u, v, p, q$ such that $p(vu)^{n\omega_M}q \in L \Leftrightarrow p(vu)^{n\omega_M}vq \notin L$ for all $n$. Let $g(r) = i \cdot p$, $h(e_1) = (vu)^{\eta\omega_{M}}$ and $h(x) = v$. We have $g(j) = g(r) \cdot h(e_1) \not \semeq{\mathcal{A}} g(r) \cdot h(e_1x) = g(k)$, showing that the pattern exists in $\mathcal{A}$. 
\end{proof}

\section{Patterns for Reverse-DFAs}\label{sec:reverse}

In this section, we formalize patterns for reverse-DFAs and show how we can move between result of DFAs and reverse-DFAs.
The following definition is almost identical to Definition \ref{def:subwordpatterns}, but for reverse deterministic automata.

\begin{definition}\label{def:reversePattern}
	Let $X$ be a set with a partial order $\preceq$. 
	A \emph{type 1 reverse subword-pattern} $\mathcal{P} = (\mathcal{S},j \neq k)$ or \emph{type 2 reverse subword-pattern} $\mathcal{P} = (\mathcal{S},j \not \leq k)$ consists of a finite partial reverse-DFA $\mathcal{S} = \left( V, X, \cdot \right)$ and two states $j,k \in V$. If $\mathcal{P} = \left( \mathcal{S}, j \neq k \right)$, we say that $\mathcal{P}$ is \emph{present} in a reverse-DFA $\mathcal{A}$ if there exists a homomorphism $h: X^* \to A^*$ where $x \preceq y$ implies that $h(x)$ is a subword of $h(y)$ and a reverse semiautomata homomorphism $g: \mathcal{S} \to \mathcal{A}^h$ such that $g(j) \not \semeq{\mathcal{A}} g(k)$ and for all $\ell \in V$, the final state of $\mathcal{A}$ can be reached from $g(\ell)$.  Analogously, we say that $\mathcal{P} = \left( \mathcal{S}, j \not \leq k \right)$ is present if there exist $h$ and $g$ such that $g(j) \not \semleq{\mathcal{A}} g(k)$. Two patterns being equivalent is defined as for non-reverse patterns, and we call the pattern \emph{reverse-rooted}\index{reverse-rooted} if there is a state $r \in V$ such that for all $\ell \in V$, we have $\ell = x \cdot r$ for some $x \in X$.
\end{definition}

Given a subword-pattern, changing the direction of the edges yields a reverse subword-pattern.

\begin{definition}
	Let $\mathcal{P} = \left( \mathcal{S}, j \not \leq k \right)$ be a subword-pattern with $\mathcal{S} = (Q,X,\cdot)$. The reverse subword-pattern $\overline{\mathcal{P}}$ is the pattern $\left( \overline{\mathcal{S}},j \not \leq k \right)$ where $\overline{\mathcal{S}} = (Q,X,\cdot^r)$ is given by $x \cdot^r \ell = \ell \cdot x$ for all $x \in X$, $\ell \in Q$.
\end{definition}

Consider $\mathcal{P}^{\mathbf{R}}_1 = (\mathcal{S}, j \neq k)$ where the underlying graph is
	
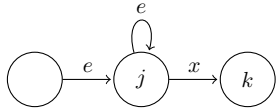
\begin{equation*}
	\begin{tikzpicture}[shorten >=1pt,node distance=1.4cm,on grid,auto,initial text = {},every node/.style={scale = 0.75}] 
		\node[state] (q_m1){$j$}; 
		\node[state] (q_m2) [left=of q_m1] {}; 
		\node[state] (q_0) [right=of q_m1] {$k$}; 
	    \path[->] (q_m1) edge[loop above]  node {$e$} (q_m1);
	    \path[->] (q_m2) edge  node {$e$} (q_m1);
	    \path[->] (q_m1) edge  node {$x$} (q_0);
	\end{tikzpicture}
  \end{equation*}
  and $x \preceq e$.
  By Corollary \ref{cor:PatternsForLevels}, this pattern characterises the \greenR-trivial languages (i.e.\ the patterns whose syntactic monoids have trivial $\greenR$-classes). The pattern $\overline{\mathcal{P}}^{\mathbf{R}}_1$ is given by the following underlying graph:
	
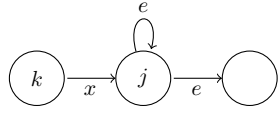
\begin{equation*}
	\begin{tikzpicture}[shorten >=1pt,node distance=1.4cm,on grid,auto,initial text = {},every node/.style={scale = 0.75}] 
		\node[state] (q_m1){$j$}; 
		\node[state] (q_m2) [right=of q_m1] {}; 
		\node[state] (q_0) [left=of q_m1] {$k$}; 
	    \path[->] (q_m1) edge[loop above]  node {$e$} (q_m1);
	    \path[<-] (q_m2) edge  node {$e$} (q_m1);
	    \path[<-] (q_m1) edge  node {$x$} (q_0);
	\end{tikzpicture}
  \end{equation*}
  Let $\mathcal{A}$ be the following reverse-DFA:
	
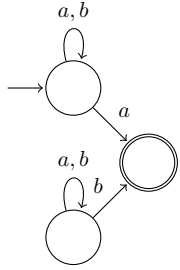
\begin{equation*}
	\begin{tikzpicture}[shorten >=1pt,node distance=1.4cm,on grid,auto,initial text = {},every node/.style={scale = 0.75}] 
		\node[state] (q_m1){}; 
		\node[state,accepting] (q_0) [above right=of q_m1] {}; 
		\node[state, initial] (q_m2) [above left=of q_0] {}; 
	    \path[->] (q_m1) edge[loop above]  node {$a,b$} (q_m1);
	    \path[->] (q_m2) edge[loop above]  node {$a,b$} (q_m2);
	    \path[->] (q_m2) edge  node {$a$} (q_0);
	    \path[->] (q_m1) edge  node {$b$} (q_0);
	\end{tikzpicture}
  \end{equation*}
  Then $L(\mathcal{A}) = A^*a$ where $A = \left\{ a,b \right\}$. We note that $\overline{\mathcal{P}}^{\mathbf{R}}_1$ is not present in $\mathcal{A}$. Since $A^*a$ is not \greenR-trivial, we see that reversing patterns does not preserve characterisation of languages. It can, however, be shown that the pattern $\overline{\mathcal{P}}^{\mathbf{R}}_1$ characterises the \greenL-trivial patterns. This hints at a left-right symmetry of the varieties characterised by a pattern $\mathcal{P}$ and $\overline{\mathcal{P}}$. We formalise this.

\begin{definition}
	Let $t \in \Omega_X$. We define the \emph{reverse $\omega$-term}\index{o-term@\omega-term!reverse}, $t^r$, inductively:
	\begin{enumerate}[(i)]
		\item If $t \in X$ or $t = 1$, then $t^r = t$,
		\item If $t = t_1t_2$ , then $t^r = t_2^rt_1^r$,
		\item If $t = s^{\omega}$ , then $t^r = (s^r)^{\omega}$.
	\end{enumerate}
	We say that an $\omega$-term $t$ is \emph{symmetric}\index{o-term@\omega-term!symmetric} if $t = t^{r}$ up to a renaming of the variables.
\end{definition}

The following lemma shows a close connection between reverse monoids and reverse $\omega$-terms.

\begin{lemma}\label{lem:WordInPreimageOfOmegaTermReverseWordInPreimageOfReverseOmegaTerm}
	Let $t \in \Omega_X$ be an omega term, and let $i : X \to M$ be a function. Then, the interpretation $I: \Omega_X \to M$ generated by $i$ satisfies $I(t) = x$ if and only if the interpretation $I': \Omega_X \to M^{r}$ generated by $i$ satisfies $I'(t^r) = x$. In particular, $M$ satisfies $t \leq s$ if and only if $M^r$ satisfies $t^r \leq s^r$, and $M$ satisfies $t = s$ if and only if $M^r$ satisfies $t^r = s^r$.
\end{lemma}

\begin{proof}
	By symmetry, we need only show that $I(t) = x$ implies $I'(t^r) = x$. We proceed by structural induction. The statement is obvious for $1$ and for variables. Next, suppose $t = t_1 t_2$. There are elements $y_1,y_2 \in M$ such that $y_1 = I(t_1)$, $y_2 = I(t_2)$ and $x = y_1y_2$. By induction, $y_1 = I'(t_1^r)$, $y_2 = I'(t_2^r)$. In $M^r$, we have $x = y_2 \cdot y_1 = I'(t_2^r) \cdot I'(t_1^r) = I'(t_2^rt_1^r) = I'(t^r)$. The case $t = s^{\omega}$ is trivial, noting that $\omega_M = \omega_{M^r}$. 
\end{proof}

\begin{definition}
	Let $\{ t_i \leq s_i \}$ be a set of $\omega$-relations and let $\mathbf{V} = \intp{t_i \leq s_i}$ be the corresponding variety, then $\mathbf{V}^r = \intp{t_i^r \leq s_i^r}$. If $\mathbf{V} = \mathbf{V}^{r}$, then $\mathbf{V}$ is \emph{symmetric}.
\end{definition}

We note, for instance that $\mathbf{R}_m = \mathbf{L}_m^r$, $\mathbf{L}_m = \mathbf{R}_m^r$ which in particular implies that the varieties $\mathbf{R}_m \cap \mathbf{L}_m$ are symmetric. The varieties $\Si_m$ are also symmetric.

\begin{lemma}\label{lem:PatternsForReverseDFASatifyReverseOmegaTerm}
	Let $\mathcal{P} = \left( \mathcal{S},j \not \leq k \right)$ be a pattern and $\overline{\mathcal{P}}$ the corresponding reverse pattern. Then $\mathcal{V} = \genby{\mathcal{P}}$ if and only if $\mathcal{V}^r = \genby{\overline{\mathcal{P}}}$. 
\end{lemma}

This Lemma shows in particular that if if $\mathbf{V}$ is symmetric, then the patterns defining membership in the corresponding language variety for a DFA is essentially the same as the ones defining it for a reverse-DFA. The only difference is the direction of the edges. In particular, this means that we have reverse-DFA characterizations of all varieties in Corollary \ref{cor:PatternsForLevels}.

\begin{proof}
	By symmetry, it is enough to show $\mathcal{V}^r = \genby{\overline{P}_i}$ implies $\mathcal{V} = \genby{P_i}$. Let $L = L(\mathcal{A})$ for some automata $\mathcal{A}$. 
	By Lemma \ref{lem:WordInPreimageOfOmegaTermReverseWordInPreimageOfReverseOmegaTerm}, we have $L \in \mathcal{V}$ if and only if $\overline{L} \in \mathcal{V}^r$. It is also clear that $L \in \genby{\mathcal{P}}$ if and only if $\overline{L} = L(\overline{\mathcal{A}}) \in \genby{\overline{\mathcal{P}}}$. We get
	\begin{equation*}
		L \in \mathcal{V} \Leftrightarrow \overline{L} \in \mathcal{V}^r \Leftrightarrow \overline{L} \in \genby{\overline{\mathcal{P}}} \Leftrightarrow L \in \genby{\mathcal{P}}.
	\end{equation*}
	Since $L$ was arbitrary, the result follows. 
\end{proof}

\section{The Finite Behaviour of Carton-Michel automata}\label{sec:PatternsForCartonMichelAutomata}

We consider two types of patterns for Carton-Michel automata, dealing with the finite and infinite behaviour respectively. To make this distinction precise, we introduce the fin-syntactic and inf-syntactic monoids. The former identifies words which behaves the same with respect to finite prefixes of the language, and the latter identifies words which behaves the same with respect to infinitely iterated words.

\begin{definition}
	Let $L \subseteq A^{\omega}$ be a language and let $u,v \in A^{*}$. We say that $u \leq_{fin} v$ if for all $x,y,z \in A^*$,
	\begin{equation*}
		xuyz^{\omega} \in L \Rightarrow xvyz^{\omega} \in L.
	\end{equation*}
	We define the \emph{fin-syntactic morphism} to be the natural projection $\pi: A^* \to \faktor{A^*}{\leq_{fin}}$ and the codomain is called the \emph{fin-syntactic monoid}.
	We define the \emph{inf-syntactic morphism} and \emph{monoid} analogously using $\leq_{inf}$ defined by $u \leq_{inf} v$ if for all $x,y \in A^*$, we have 
	\begin{equation*}
		x(uy)^{\omega} \in L \Rightarrow x(vy)^{\omega} \in L.
	\end{equation*}
\end{definition}

It is clear that the syntactic semigroup is in some variety $\mathbf{V}$ if and only if both the fin-syntactic monoid and inf-syntactic semigroup are in $\mathbf{V}$.

In this section, we deal with the behaviour of the fin-syntactic monoid. This is done using patterns which are defined almost exactly as Definition \ref{def:reversePattern}. However, one needs to be careful in the definition of such a pattern being present. In fact, since Carton-Michel automata are not guaranteed to be reverse-deterministic everywhere, we can not a priori define \emph{any} candidate witness $g: \mathcal{S} \to \mathcal{A}$. However, this is easily remedied by considering morphisms to \emph{trim} components of the Carton-Michel automata.

\begin{definition}\label{def:CMApattern}
	A \emph{type 1} and \emph{type 2 subword-pattern} for a Carton-Michel automata is defined as in Definition \ref{def:reversePattern}. It is \emph{present} in a Carton-Michel automaton $\mathcal{A}$ if there exists a homomorphisms $h: X^* \to A^*$, a trim subautomaton $\mathcal{B}$ in $\mathcal{A}$, and a reverse semiautomata homomorphism $g: \mathcal{S} \to \mathcal{B}^h$ where $g$ and $h$ has the properties of Definition \ref{def:reversePattern}.
\end{definition}

Note that it is not sufficient to assume that every state in the automata is reachable from a final state. Consider for instance the following automata recognizing the language $A^*aA^{\omega}$. This language has syntactic monoid in $\Si_1 = \mathbf{J}^+$. However, both $k_4$ and $k_5$ are reachable from a final state, making $g(k) = k_4$ and $g(j) = k_5$ a viable witness for $\overline{\mathcal{P}^{\Si}_1}$ being present if only reachability from final states was required.
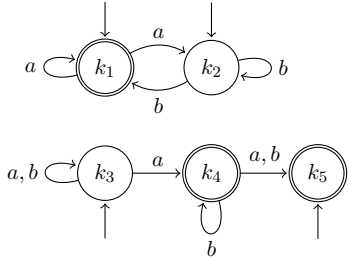
\begin{equation*}
	\begin{tikzpicture}[shorten >=1pt,node distance=1.4cm,on grid,auto,initial text = {},every node/.style={scale = 0.75}] 
		   \node[state, accepting, initial above] (q1) {$k_1$}; 
		   \node[state, initial above] (q2) [right=of q1] {$k_2$}; 
		   \node[state, initial below] (q3) [below=of q1] {$k_3$}; 
		   \node[state, accepting] (q4) [right=of q3] {$k_4$}; 
		   \node[state, accepting, initial below] (q5) [right=of q4] {$k_5$}; 
		    \path[->] (q1) edge[loop left]  node {$a$} (q1);
		    \path[->] (q2) edge[loop right]  node {$b$} (q2);
		    \path[->] (q1) edge[bend left]  node {$a$} (q2);
		    \path[->] (q2) edge[bend left]  node {$b$} (q1);
		    \path[->] (q3) edge[loop left]  node {$a,b$} (q3);
		    \path[->] (q4) edge[loop below]  node {$b$} (q4);
		    \path[->] (q3) edge  node {$a$} (q4);
		    \path[->] (q4) edge  node {$a,b$} (q5);
		\end{tikzpicture}
	\end{equation*}

The following lemma show that $M_{fin} \in \mathbf{V}$ can be characterised by using the same patterns as in the finite reverse-DFA case.

\begin{lemma}\label{lem:MfinInVEquivAdoesNotHavePattern}
	Let $\mathcal{A}$ be a Carton-Michel automaton, recognising a language $L(\mathcal{A})$ with fin-syntactic morphism $\mu: A^* \to M_{fin}$. Let $\mathcal{V} = \genby{ \mathbf{P}}$ where the patterns are language patterns for reverse-DFAs. Then $M_{fin} \in \mathbf{V}$ if and only if $\mathcal{A}$ does not have any of the patterns in $\mathbf{P}$.
\end{lemma}

\begin{proof}
	Since a pattern is present if and only if it is present in the trim subautomata $\mathcal{A}$, we lose no generality in assuming $\mathcal{A}$ to be trim.
	For each $\ell \in Q$, consider the reverse-DFA $\mathcal{A}_\ell = \left( Q,A,\cdot, I, \left\{ \ell \right\} \right)$. The syntactic morphism $\mu_{\ell} : A^* \to M_\ell = \faktor{A^*}{\leq_\ell}$ of $L(\mathcal{A}_{\ell})$ is given by the natural projection on the equivalence classes of the relation $\leq_{\ell}$ defined by $u \leq_{\ell} v$ if
	\begin{equation*}
		xuy \cdot \ell \in I \Rightarrow xvy \cdot \ell \in I
	\end{equation*}
	for all $x,y \in A^*$. We have $M_\ell \in \mathbf{V}$ if and only if $\mathcal{A}_\ell$ does not have any of the patterns in $\mathbf{P}$.

	Suppose $\mu(u) \leq \mu(v)$ and $\ell = \startof{y'z^{\omega}}$. Since $xuyy'z^{\omega} \in L$ implies $xvyy'z^{\omega} \in L$ for all $x,y \in A^*$, we get $\mu_\ell(u) \leq \mu_\ell(v)$. Thus $M_{\ell}$ divides $M$. Since $\mathcal{A}$ is trim, this is true for all states $\ell$ in $\mathcal{A}$ and it follows that if $\mathcal{A}$ has a pattern $\mathcal{P} \in \mathbf{P}$, then so does some $\mathcal{A}_{\ell}$, and thus $M \notin \mathbf{V}$.

	On the other hand, if $\mu_\ell(u) \leq \mu_\ell(v)$ for all $\ell$, then $xuyz^{\omega} \in L$ implies $xvyz^{\omega} \in L$ for all $z$. Hence $M$ divides $M_{\ell_1} \times \dots \times M_{\ell_k}$. It follows that if $M_\ell \in \mathbf{V}$ for all $\ell$, then so is $M$. If $M_{\ell} \in \mathbf{V}$ for all $\ell$, then no $\mathcal{A}_{\ell}$ has any of the patterns in $\mathbf{P}$ which implies that $\mathcal{A}$ can not have any of the patterns. 
\end{proof}

\section{The Infinite Behaviour of Carton-Michel automata}\label{sec:InfBehaviour}

In this section, we give pattern characterizations of the infinite behaviour of Carton-Michel automata. We characterize two types of infinite behaviour. First, we handle the inf-syntactic monoid, and show that for our purposes it is enough to show that it is in $\DA$. Then, we give pattern characterizations for being open, closed respectively clopen in the Cantor and alphabetic topology.

We use a modified version of subword-patterns, \emph{enhanced subword-patterns}. The enhancement is twofold; we assume that every path corresponding to an edge in the pattern is non-empty, and we assume that some edges can be distinguished as \emph{final}. The paths corresponding to these edges are required to have some final state along them.

\begin{definition}\label{def:enhancedPattern}
	Let $X$ be a set with a partial order $\preceq$. 
	A \emph{type 1 reverse subword-pattern} $\mathcal{P} = (\mathcal{S},j \neq k, F)$ or \emph{type 2 reverse subword-pattern} $\mathcal{P} = (\mathcal{S},j \not \leq k, F)$ are defined as in Definition \ref{def:CMApattern} with $F$ being a subset of the edges in $\mathcal{S}$ called \emph{final edges}.
	If $\mathcal{P} = \left( \mathcal{S}, j \neq k \right)$, we say that $\mathcal{P}$ is \emph{present} in a Carton-Michel automata $\mathcal{A}$ if there exists a homomorphism $h: X^+ \to A^+$ where $x \preceq y$ implies that $h(x)$ is a subword of $h(y)$ and a trim subautomata $\mathcal{B}$ with a reverse semiautomata homomorphism $g: \mathcal{S} \to \mathcal{B}^h$ such that $g(j) \not \semeq{\mathcal{A}} g(k)$ with the following property: if $x \in F$, $x \cdot \ell$ is defined and $h(x) = a_1 \dots a_n$, then there exists $1 \leq i \leq n$ such that $a_i \dots a_n \cdot g(\ell)$ is final.
\end{definition}

We show that for varieties $\mathbf{J}_1 \subseteq \mathbf{V} \subseteq \DA$, characterizing having syntactic monoid in $\mathbf{V}$ is equivalent to having fin-syntactic monoid in $\mathbf{V}$ and inf-syntactic monoid in $\DA$. We also handle the special case $\Si_1 = \mathbf{J}^+$ which does not include $\mathbf{J}_1$.

\begin{lemma}\label{lem:DecideFO2HiearchiesViaFinsemigroup}
	Let $\mathbf{V}$ be a (positive) variety such that $\mathbf{J}_1 \subseteq \mathbf{V} \subseteq \DA$. Let $L \subseteq A^{\infty}$ be a language with syntactic monoid $M$, fin-syntactic morphism $\mu_{fin}: A^* \to M_{fin}$ and inf-syntactic morphism $\mu_{inf}: A^* \to M_{inf}$. Then $M \in \mathbf{V}$ if and only if $M_{fin} \in \mathbf{V}$ and $M_{inf} \in \DA$.
\end{lemma}

\begin{proof}
	It is clear that if $M \in \mathbf{V}$, then so are $M_{fin}$ and $M_{inf}$. 
	For the other direction, assume $M_{fin} \in \mathbf{V}$ and $M_{inf} \in \DA$. We note that the (unordered) monoid $2^A$ with union as operation is in $\mathbf{J}_1$. Let $M'$ be the submonoid of $M_{fin} \times 2^A$ generated by $\nu(a) = (\mu_{fin}(a),\left\{ a \right\})$. We show that there exists a surjective homomorphism $f: M' \to M_{inf}$. This implies $M_{inf} \in \mathbf{V}$ which implies $M \in \mathbf{V}$.

	Let $u \in A^+$ and define $f\left( \nu(u) \right) = \mu_{inf}(u)$. We need to show that this is well-defined. In other words, we need to show that $\nu(u) \leq \nu(v)$ implies $\mu_{inf}(u) \leq \mu_{inf}(v)$. Let us assume the former, that is, we assume $\mu_{fin}(u) \leq \mu_{fin}(v)$ and $\alp(u) = \alp(v)$. We set $n = \omega_M$, and get
  \begin{alignat}{2}
	  x(uy)^{\omega} = x\left((uy)^n  \right)^{\omega} \in L 
	  \quad \Rightarrow \quad & x\left( (uy)^n(vy)^{2n}(uy)^n \right)^{\omega} \in L \label{aaa:FinSemigroupProof}
	  \\ \Rightarrow \quad & x(uy)^n(vy)^n\left( (vy)^n(uy)^{2n}(vy)^n \right)^{\omega} \in L \label{bbb:FinSemigroupProof}
	  \\ \Rightarrow \quad & x(uy)^n(vy)^{\omega}\in L \label{ccc:FinSemigroupProof}
	  \\ \Rightarrow \quad & x(vy)^n(vy)^{\omega} = x(vy)^{\omega} \in L \label{ddd:FinSemigroupProof}
  \end{alignat}
  where (\ref{aaa:FinSemigroupProof}) and (\ref{ccc:FinSemigroupProof}) follows from $M_{inf} \in \DA$ and $\alp(u) = \alp(v)$, (\ref{ddd:FinSemigroupProof}) follows from $\mu_{fin}(u) \leq \mu_{fin}(v)$ and (\ref{bbb:FinSemigroupProof}) is just a rewriting of the word. 
\end{proof}

Out of the monoids appearing in Table \ref{tbl:MonoidCriteria}, only $\Si_1$ does not contain $\mathbf{J}_1$. Thus, we only need to find two pattern characterizations, one for the inf-syntactic monoid being in $\Si_1$ and one for it being in $\DA$. For the latter, the following lemma is useful.

\begin{lemma}\label{lem:AssumeYIsEmpty}
	Let $n \in \mathbb{N}$ be fixed, and let $L \subseteq A^{\omega}$ be a language. Let $\mu_{fin} : A^* \to M_{fin}$ be its fin-syntactic morphism and let $M_{inf}$ be its inf-syntactic monoid. If $M_{fin} \in \DA$ and $M_{inf} \notin \DA$, then there exists $x \in A^*$ $e,u \in A^+$ with $\alp(u) \subseteq \alp(e)$ such that $x(e^{n}ue^n)^{\omega} \in L \Leftrightarrow x(e^n)^{\omega} \notin L$ for all $n$.
\end{lemma}

\begin{proof}
	Let $k = \omega_{M_{inf}}\omega_{M_{fin}}$. It follows directly from the fact that $M_{inf} \notin \DA$ that there exists $x,y \in A^*$, $e,u \in A^+$ with $\alp(u) \subseteq \alp(e)$ such that $x(e^{k}ue^ky)^{\omega} \in L$ while $x(e^ky)^{\omega} \notin L$.

	Let $f_1 = (e^kue^ky)^k$ and $f_2 = (e^ky)^k$. Note that $\alp(f_1) = \alp(f_2)$. We have $x(f_1^nf_2^nf_1^n)^{\omega} \in L \Leftrightarrow xf_1^n(f_2^nf_1^n)^{\omega} \in L \Leftrightarrow xf_2^n(f_2^nf_1^n)^{\omega} \in L \Leftrightarrow x(f_2^nf_1^nf_2^n)^{\omega} \in L$ where the middle equivalence follows from the fact that $\mu_{fin}(f_1) = \mu_{fin}(f_2)$. Thus, it must be the case that either $x(f_1^n)^{\omega} \in L \not \Leftrightarrow x(f_1^nf_2^nf_1^n)^{\omega} \in L$ or $x(f_2^n)^{\omega} \in L \not \Leftrightarrow x(f_2^nf_1^nf_2^n)^{\omega} \in L$. This gives the desired result. 
\end{proof}

\begin{proposition}\label{prp:MinfInDACanBeCheckedByPattern}
	Let $\mathcal{A}$ be a Carton-Michel automaton, and let $M_{inf}$ be the inf-syntactic monoid of $L(\mathcal{A})$. Let $\mathcal{S}_{si}$ and $\mathcal{S}_{da}$ be the following partial semiautomata:
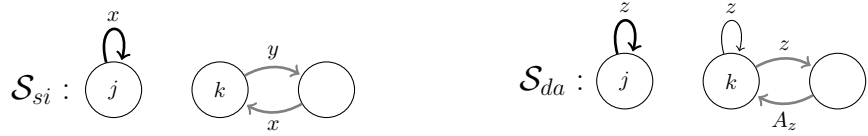
\begin{equation*}
	\begin{tikzpicture}[shorten >=1pt,node distance=1.4cm,on grid,auto,initial text = {},every node/.style={scale = 0.75}] 
	   \node[state] (q_0) {$k$}; 
	   \node[state] (q_1) [right=of q_0] {}; 
	   \node[state] (q_2) [left=of q_0] {$j$}; 
	\node [left=of q_2] {\qquad\quad\LARGE$\mathcal{S}_{si}:$};
	    \path[->] (q_1) edge[bend left,line width=1,gray]  node {\color{black}$x$} (q_0);
	    \path[->] (q_0) edge[bend left,line width=1,gray]  node {\color{black}$y$} (q_1);
	    \path[->] (q_2) edge[loop above,line width=1]  node {$x$} (q_2);
	\end{tikzpicture}\qquad\quad
	\begin{tikzpicture}[shorten >=1pt,node distance=1.4cm,on grid,auto,initial text = {},every node/.style={scale = 0.75}] 
	   \node[state] (q_0) {$k$}; 
	   \node[state] (q_1) [right=of q_0] {}; 
	   \node[state] (q_2) [left=of q_0] {$j$}; 
	\node [left=of q_2] {\qquad\quad\LARGE$\mathcal{S}_{da}:$};
	    \path[->] (q_0) edge[loop above]  node {$z$} (q_0);
	    \path[->] (q_1) edge[bend left,line width=1,gray]  node {\color{black}$A_z$} (q_0);
	    \path[->] (q_0) edge[bend left,line width=1,gray]  node {\color{black}$z$} (q_1);
	    \path[->] (q_2) edge[loop above,line width=1]  node {$z$} (q_2);
	\end{tikzpicture}
\end{equation*}
where $A_z \preceq z$ and for each pattern the black bold edge as well as at least one of the gray bold edges are final edges. We then have the following characterizations:
\begin{enumerate}[(i)]
	\item $M_{inf} \in \Si_1$ if and only if $\mathcal{P}^{\Si}_{1\text{-}inf} = (\mathcal{S}_{si},j \not \leq k)$ is not present in $\mathcal{A}$,\label{aaa:MinfInDACanBeCheckedByPattern}
	\item suppose $\overline{\mathcal{P}}_{\DA}$ is not in $\mathcal{A}$, then $M_{inf} \in \DA$ if and only if $\mathcal{P}_{\DA\text{-}inf} = (\mathcal{S}_{da},j \neq k)$ is not present in $\mathcal{A}$,\label{bbb:MinfInDACanBeCheckedByPattern}
\end{enumerate}
\end{proposition}

\begin{proof}
	Let $\mu_{inf}: A^* \to M_{inf}$ be the inf-syntactic morphism.
	Let us first consider \itref{aaa:MinfInDACanBeCheckedByPattern}. Assume that the pattern is present with $h: X^+ \to A^+$ as a witness. Then there exists $p$ such that $ph(x)^{\omega} \in L$ while $p(h(y)h(x))^{\omega} \notin L$. In particular, $1 \not \leq \mu_{inf}(h(y))$, showing that $M \notin \Si_1$. On the other hand, if $M_{inf} \notin \Si_1$, then there exists $p,u,v \in A^*$ such that $pu^{\omega} \in L$ but $p(vu)^{\omega} \notin L$. Defining $h(x) = u$, $h(y) = v$, $g(j) = \startof{u^{\omega}}$ and $g(k) = \startof{(vu)^{\omega}}$ gives the desired witness.

	Next, consider \itref{bbb:MinfInDACanBeCheckedByPattern}. Assume that the pattern is present with $h: X^+ \to A^+$ as a witness. Then there exists $p$ such that $p(h(z)^{n}h(A_z)h(z)^{n})^{\omega} \in L \Leftrightarrow p(h(z)^{n})^{\omega} \notin L$ for all $n$. In particular, $\mu(h(z))^{\omega_M} \mu(h(A_z)) \mu(h(z))^{\omega_M} \neq \mu(h(z))^{\omega_M}$. Since $\alp(h(A_z)) \subseteq \alp(h(z))$, Lemma \ref{lem:DAproperty} implies that $M_{inf} \notin \DA$.

	For the other direction, suppose that $M_{inf} \notin \DA$. By Lemma \ref{lem:AssumeYIsEmpty}, there exists $x \in A^*$, $e,u \in A^+$ with $\alp(u) \subseteq \alp(e)$ such that $x(e^{\eta}ue^{\eta})^{\omega} \in L \Leftrightarrow x(e^{\eta})^{\omega} \notin L$. Letting $g(j) = \startof (e^{\eta})^{\omega}$, $g(k) = \startof (e^{\eta}ue^{\eta})^{\omega}$, $h(z) = e^{|u|\eta}$ and $h(A_z) = u$ shows that $\mathcal{P}_{\DA\text{-}inf}$ is present in $\mathcal{A}$. 
\end{proof}

This leads to the following theorem characterizing membership of the inf-syntactic monoid in the varieties which interests us in this paper.

\begin{theorem}\label{thm:Characterization}
	Let $\mathcal{A}$ be a Carton-Michel automaton, and let $M$ be the syntactic monoid of $L(\mathcal{A})$. Then for $m \geq 2$:
	\begin{enumerate}[(i)]
		\item $M \in \mathbf{Si}_1$ if and only if neither $\overline{\mathcal{P}}^{\Si}_1$ nor $\mathcal{P}^{\Si}_{1\text{-}inf}$ is present in $\mathcal{A}$,\label{aaa:Characterization}
		\item $M \in \Si_m$ if and only if neither $\overline{\mathcal{P}}^{\Si}_m$, $\overline{\mathcal{P}}_{\DA}$ nor $\mathcal{P}_{\DA\text{-}inf}$ is present in $\mathcal{A}$,\label{bbb:Characterization}
		\item $M \in \mathbf{R}_{m} \cap \mathbf{L}_{m}$ if and only if neither $\overline{\mathcal{P}}^{\mathbf{R}}_{m}$,$\overline{\mathcal{P}}^{\mathbf{L}}_{m}$ nor $\mathcal{P}_{\DA\text{-}inf}$ is present in $\mathcal{A}$.\label{ccc:Characterization}
	\end{enumerate}
\end{theorem}

Next, we turn to characterizing topology. We consider patterns for the Cantor and alphabetic topology. One can obtain patterns for being closed in the respective topology by switching $j$ and $k$, and for being clopen (i.e. both open and closed) by replacing the inequality by an equality. 

\begin{proposition}\label{prp:CantorOpenCanBeCheckedByPattern}
	Let $\mathcal{A}$ be a Carton-Michel automaton, and let $\mathcal{S}_c$ and $\mathcal{S}_a$ be the partial semiautomata defined below:
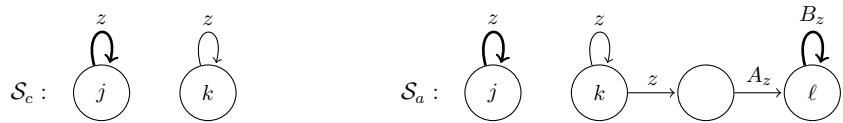
\begin{equation*}
	\begin{tikzpicture}[shorten >=1pt,node distance=1.4cm,on grid,auto,initial text = {},every node/.style={scale = 0.75}] 
		\node [left=of q_2] {\qquad\quad\large$\mathcal{S}_c:$};
		\node[state] (q_0) {$k$}; 
		\node[state] (q_2) [left=of q_0] {$j$}; 
	    \path[->] (q_0) edge[loop above]  node {$z$} (q_0);
	    \path[->] (q_2) edge[loop above,line width = 1]  node {$z$} (q_2);
	\end{tikzpicture}\qquad\quad
	\begin{tikzpicture}[shorten >=1pt,node distance=1.4cm,on grid,auto,initial text = {},every node/.style={scale = 0.75}] 
		\node [left=of q_2] {\qquad\quad\large$\mathcal{S}_a:$};
	   \node[state] (q_0) {$k$}; 
	   \node[state] (q_1) [right=of q_0] {}; 
	   \node[state] (q_3) [right=of q_1] {$\ell$}; 
	   \node[state] (q_2) [left=of q_0] {$j$}; 
	    \path[->] (q_3) edge[loop above,line width=1]  node {$B_z$} (q_3);
	    \path[->] (q_0) edge[loop above]  node {$z$} (q_0);
	    \path[->] (q_0) edge  node {$z$} (q_1);
	    \path[->] (q_1) edge  node {$A_z$} (q_3);
	    \path[->] (q_2) edge[loop above,line width=1]  node {$z$} (q_2);
	\end{tikzpicture}
\end{equation*}
Where for $\mathcal{S}_a$, we have $A_z,B_z \preceq z$. Then
\begin{enumerate}[(i)]
	\item $L(\mathcal{A}) \in \mathcal{O}_{cantor}$ if and only if $\mathcal{P}_{cantor} = (\mathcal{S}_c,j \not\leq k)$ is not present in $\mathcal{A}$,\label{aaa:CantorOpenCanBeCheckedByPattern}
	\item $L(\mathcal{A}) \in \mathcal{O}_{alph}$ if and only if $\mathcal{P}_{alph} = (\mathcal{S}_a,j \not\leq k)$ is not present in $\mathcal{A}$,\label{bbb:CantorOpenCanBeCheckedByPattern}
	\item $L(\mathcal{A})$ is clopen in the alphabetic topology if and only if $\mathcal{P}_{alph\text{-}clopen} = (\mathcal{S}_a,k \neq j)$ is not present in $\mathcal{A}$.\label{ccc:CantorOpenCanBeCheckedByPattern}
\end{enumerate}
\end{proposition}

\begin{proof}
	Let $\mu: A^* \to M$ be the syntactic morphism of $L(\mathcal{A})$. We first show \itref{aaa:CantorOpenCanBeCheckedByPattern}.
	Suppose $\mathcal{P}_{cantor}$ exists in $\mathcal{A}$ with $h(z) = u$. Choose $\alpha \in A^{\omega}$ such that $g(k) = \startof \alpha$ (such an $\alpha$ exists since $g(k)$ is reachable from some cycle with a final state). Then there exists $p$ such that $pu^{\omega} \in L$ while $pu^{n}\alpha \notin L$ for any $n$. This means $L(\mathcal{A}) \notin \mathcal{O}_{cantor}$.

	On the other hand, suppose $L(\mathcal{A}) \notin \mathcal{O}_{cantor}$. Then there exists a linked pair $(s,f)$ in $M$ and an idempotent $f'$ such that $[s][f]^{\omega} \subseteq L$ and $[s][f']^{\omega} \cap L(\mathcal{A}) = \emptyset$. Let $p \in [s]$, $u \in [f]$ and $v \in [f']$. We choose $h(z) = u^{\eta}$, $g(j) = \startof u^{\omega}$ and $g(k) = \startof u^{\eta}v^{\omega}$. Since $\mu(su^{\eta}) = \mu(s)$, it follows that $p \cdot g(j) \in I$ while $p \cdot g(k) \notin I$ giving the desired pattern.

	Showing \itref{bbb:CantorOpenCanBeCheckedByPattern} follow a similar line of argument. 
	Suppose $\mathcal{P}_{alph}$ exists with $h: Y^{+} \to A^+$ as witness. Then there is a word $x$ such that $xh(z)^{\omega} \in L$ while $xh(z)^{n}h(A_z)h(B_z)^{\omega} \notin L$ for all $n$. Since $\alp(h(A_z)), \alp(h(B_z)) \subseteq \alp(h(z))$, it follows that $L$ is not open in the alphabetic topology.

	For the other direction, suppose $L$ is not open in the alphabetic topology. Then there exists $\alpha \in L$ and $n \in \mathbb{N}$ such that $p\im(\alpha)^{\omega} \subseteq L$ for no prefix $p$ of $\alpha$ which has length at least $n$. Let $\mu: A^* \to M$ be the syntactic morphism, and let $(s,f)$ be a linked pair such that $\alpha \in [s][f]^{\omega}$. By the choice of $\alpha$, there exists $\beta = \hat{s}\hat{x}\hat{f'}^{\omega} \notin L$ such that $\mu(\hat{s}) = s$ and $\alp(\hat{f'}), \alp(\hat{x}) \subseteq \alp(\hat{f})$ where $\hat{f} \in [f]$. By concatenating with $\hat{f}$ if necessary, we can assume $\hat{x}$ is nonempty. Set $h(z) = \hat{f}^{\eta|\hat{x}\hat{f'}|}$, $h(A_z) = \hat{x}$, $h(B_z) = \hat{f'}^{\eta}$, then $g(j) = \startof \hat{f}^{\omega}$, $g(\ell) = \startof{\hat{f'}^{\omega}}$ gives the desired pattern.

	By considering the complement of the language, we see that whenever a pattern $\mathcal{P} = \left( \mathcal{S}, j \not \leq k \right)$ characterizes being open a topology, then $\mathcal{P} = \left( \mathcal{S}, k \not \leq j \right)$ characterizes being closed and thus $\mathcal{P} = \left( \mathcal{S}, k \neq j \right)$ characterizes being clopen. Thus \itref{bbb:CantorOpenCanBeCheckedByPattern} implies \itref{ccc:CantorOpenCanBeCheckedByPattern}. 
\end{proof}

To explicitly mention clopen-ness in the alphabetic topology has a purpose. The following two lemmas show that an $\omega$-language with syntactic monoid in $\DA$ is in the Boolean closure of the Cantor topology if and only if it is clopen in the alphabetic topology.

\begin{lemma}\label{lem:ClopenInAlphImpliesCantorBool}
	If a regular language $L \subseteq A^{\omega}$ satisfy $L \in \mathcal{O}_{alph}$ and $A^{\omega} \setminus L \in \mathcal{O}_{alph}$ then it is in $\mathbb{B}(\mathcal{O}_{cantor})$.
\end{lemma}

\begin{proof}[Lemma \ref{lem:ClopenInAlphImpliesCantorBool}]
	Let $\mu: A^* \to M$ be the syntactic morphism of $L$. We  show that for every linked pair $(s,e)$ in $M$, there exists $L' \in \mathbb{B}(\mathcal{O}_{cantor})$ such that either $[s][e]^{\omega} \subseteq L' \subseteq L$ or $[s][e] \subseteq L' \subseteq A^{\omega} \setminus L$. This implies the desired result.
	
	We consider the directed graph $\mathcal{F}$ with vertices all linked pairs $(t,f)$ and an edge $(t,f) \to (t',f')$ if and only if $tp = t'$ for some $p$ and $[t][f] \subseteq L \Leftrightarrow [t'][f'] \subseteq A^{\omega} \backslash L$. We show that $\mathcal{F}$ is a forest. 
	Indeed, suppose $tp = t'$ and $t'q = t$. Then $tfpf'q = t$ and thus $(t,(fpf'q)^{\omega_M})$ is a linked pair. Without loss of generality, assume $[t][(fpf'q)^{\omega_M}]^{\omega} \subseteq L$. It follows by alphabetic openness that $[t][f]^{\omega} \subseteq L$ and $[t'][f']^{\omega} \subseteq L$, a contradiction. Since $\mathcal{F}$ is a forest, we can define a well order $(t,f) \leq_{\mathcal{F}} (t',f')$ if and only if $(t,f)$ is reachable from $(t',f')$.

	We use induction over $\leq_{\mathcal{F}}$.
	By symmetry, it is enough to show that the desired $L'$ exists for $[s][e]^{\omega} \subseteq L$.
	We have that $[s][e]^{\omega} \subseteq [s]A^* \setminus \bigcup [t][f]^{\omega} \subseteq L$ where the union is taken over all $(t,f) <_{\mathcal{F}} (s,e)$ such that $[t][f]^{\omega} \cap L = \emptyset$. By induction, for each such pair, there is a set $X_{t,f} \in \mathbb{B}(\mathcal{O}_{cantor})$ such that $X_{t,f} \cap L = \emptyset$. We get $[s][e]^{\omega} \subseteq [s]A^* \setminus \bigcup X_{t,f} \subseteq L$ which yields the desired result. 
\end{proof}

The other direction is not true in general. In particular, every singleton $\left\{ \alpha \right\}$ is closed in the Cantor topology. If it was open in the alphabetic topology, then any language would be. However, we have the following special case. 

\begin{lemma}\label{lem:IfFinDAThenCantorBoolImpliesOpenAlph}
	If $\mathcal{A}$ is a Carton-Michel automaton in which $\overline{\mathcal{P}}_{\DA}$ is not present, and if $L(\mathcal{A}) \in \mathbb{B}(\mathcal{O}_{cantor})$, then $L(\mathcal{A}) \in \mathcal{O}_{alph}$ and $L(\mathcal{A}) \in A^{\omega} \setminus \mathcal{O}_{alph}$.
\end{lemma}

\begin{proof}
	We show that $\mathbb{B}(\mathcal{O}_{cantor}) \subseteq \mathcal{O}_{alph}$ which by symmetry implies the result. If $L \in \mathcal{O}_{cantor}$, then it is clearly in $\mathcal{O}_{alph}$.

	Next, assume $A^{\omega} \setminus L \in \mathcal{O}_{cantor}$.
	Let $u \in A^*$ and suppose $pu^{\omega} \in L$. We want to show $pu^{\eta}v^{\omega} \in L$ for all $v$ with $\alp(v) \subseteq \alp(u)$ which would yield the desired result. For contradiction, assume $pu^{\eta}v^{\omega} \notin L$. Since $A^{\omega} \setminus L \in \mathcal{O}_{cantor}$ we have $n$ such that $pu^{\eta}v^nA^{\omega} \cap L(\mathcal{A}) = \emptyset$. In particular $pu^{\eta}v^nu^{\omega} \notin L$. Setting $g(j) = \startof u^{\eta}v^nu^{\omega}$, $g(k) = \startof u^{\omega}$, $h(x) = u^{n|v|\eta}$ and $h(A_x) = u^{\eta}v^n$ gives witnesses of the pattern $\overline{\mathcal{P}}_{\DA}$ being present in $\mathcal{A}$, a contradiction. 
\end{proof}

To conclude this contribution, we note some optimizations; for deciding membership in $\FO^2_1$, $\Sigma^2_1$ and $\Sigma^2_2$, the patterns $\mathcal{P}^{\Si}_{1\text{-}inf}$ and $\mathcal{P}_{\DA\text{-}inf}$ are redundant.

\begin{lemma}\label{lem:CantorOpenImpliesInfSiOne}
	Let $\mathcal{A}$ be a Carton-Michel automaton in which $\overline{\mathcal{P}}^{\Si}_{1}$ is not present. If $\mathcal{P}^{\Si}_{1\text{-}inf} = \left( \mathcal{S},j \not \leq k \right)$ is present in $\mathcal{A}$, then so is $\mathcal{P}_{cantor}$.
\end{lemma}

\begin{proof}
	Suppose $h, g$ are witnesses for $\mathcal{P}^{\Si}_{1\text{-}inf}$ being present in $\mathcal{A}$. Let $\ell_n = h(x)^{n} \cdot g(k)$. We show $\ell_{n} \semeq{\mathcal{A}} g(k) \not \semgeq{\mathcal{A}} g(j)$, which gives the desired result since $h(x)^{\eta}$ is a loop at both $g(j)$ and $\ell_{\eta}$.

	We use induction over $n$. For $n = 0$, there is nothing to show. Now suppose $\ell' = h(x) \cdot g(k)$, and $g(k) = h(y) \cdot \ell'$. Since $\mathcal{P}^{\Si}_{1}$ is not present, we have $\ell' \semeq{\mathcal{A}} g(j)$. We get
  \begin{alignat*}{3}
	  ph(x)^{n+1}g(j) \in I \quad 
	  \Leftrightarrow \quad & ph(x)^{n}\cdot \ell'  \in I && \quad \text{}
	  \\ \Leftrightarrow \quad & ph(x)^{n}\cdot g(j)  \in I && \quad \text{since $\ell' \semeq{\mathcal{A}} g(j)$.}
	  \\ \Leftrightarrow \quad & p \cdot g(j)  \in I && \quad \text{by induction,}
  \end{alignat*}	
  showing that $\ell_n \semeq{\mathcal{A}} g(j)$. 
\end{proof}

\begin{lemma}\label{lem:AlpOpenImpliesInDA}
	Let $\mathcal{A}$ be a Carton-Michel automaton, and suppose that $\mathcal{A}$ has the pattern $\mathcal{P}_{\DA\text{-}inf}$, then $\mathcal{A}$ has the pattern $\mathcal{P}_{alph}$ or the pattern $\overline{\mathcal{P}}_{\DA}$.
\end{lemma}

\begin{proof}
	To differentiate the variables from the two different patterns, we prime all variables used in $\mathcal{P}_{alph}$. We first assume that $\mathcal{P}_{\DA\text{-}inf}$ exists in $\mathcal{A}$ witnessed by the automata morphism $g$ and the monoid homomorphism $h$ where $g(j) \not \semleq{\mathcal{A}} g(k)$. We define $h'$ and $g'$ witnessing the existence of the pattern $\mathcal{P}_{alph}$ by 
	\begin{align*}
		g'(k') & = g'(\ell') = g(k),
		& g'(j') & = g(j),
		\\ h'(z) & = h(z)^2
		& h'(A_z') &= h(z),
		& h'(B_z) & = h(zA_z).
	\end{align*}
	Since $h(A_z)$ is a subword of $h(z)$, it follows that $h'(A_{z}')$ is a subword of $h'(z)$, and thus $h'$, $g'$ witnesses the desired pattern.

	On the other hand, if $g(j) \not \semgeq{\mathcal{A}} g(k)$, we define 
	\begin{align*}
		g'(\ell') & = g(j),
		& g'(k') & = h(z^{\eta}A_z)^{\eta} \cdot g(j),
		& g'(j') & = g(k),
		\\ h'(z') & = h(z^{\eta}A_z)^{\eta}
		& h'(A_z') &= h'(B_z') = h(z).
	\end{align*}
	If $g'(k') \not \semeq{\mathcal{A}} g(j)$, then since $h(A_z)$ is a subword of $h(z^{\eta})$, we have that $\overline{\mathcal{P}}_{\DA}$ is present in $\mathcal{A}$. On the other hand, if $g'(k') \semeq{\mathcal{A}} g(j)$, then $g'(j') = g(k) \not \semleq{\mathcal{A}} g(j) \semeq{\mathcal{A}} g'(k')$, and thus the pattern $\mathcal{P}_{alph}$ is present. 
\end{proof}

\section{$\NL$-completeness}
\label{sec:Complexity}

In general, deciding membership on DFA-input is intractable (e.g.\ deciding membership of star-free languages is $\mathbf{PSPACE}$-complete \cite{ChoHuynh91tcs}). 
However, one of the advantages of using patterns to characterize some variety is that deciding the presence of patterns is in $\NL$ and thus so is variety membership. We show that this is also true for subword-patterns which have \emph{stable superwords}.

\begin{definition}
	Let $\mathcal{P}$ be a pattern such that whenever $x \preceq y$ such that $x \neq y$, and $\ell \circ y$ is defined, then $\ell \circ yy = \ell \circ y$. We say that $\mathcal{P}$ has the \emph{stable superwords}.
\end{definition}

Intuitively, a pattern has the stable superwords if all $y$ which are not minimal with respect to $\preceq$ occurs only as transitions to a state where $y$ is a loop. Note in particular that all patterns which we have introduced explicitly throughout the paper has this property. We also note that if $\preceq$ is the identity, then the pattern vacuously has stable superwords.

We introduce Algorithm \ref{alg:subwordpattern} for patterns having stable superwords. It finds a given subword-pattern $\mathcal{P}$ non-deterministically storing only a finite number of states (depending on $\mathcal{P}$) from the automata, thus using only logarithmic space. The fundamental idea of the algorithm is as in the non-subword case (cf.~\cite{ChoHuynh91tcs,GlasserSchmitz08tocs}); for each variable in the pattern, we trace out paths in the automata, remembering only the initial and final states. However, we need to take special care for variables $x \preceq y$. Every time we want to take a step on the path of $x$, we require that we also take a step on the path of $y$, showing the desired subword property.

\begin{algorithm}[h]
\DontPrintSemicolon
\SetAlgoLined
\KwData{A DFA $\mathcal{A} = \left( Q,A,i,E,\cdot \right)$}
\ForEach{$\ell \in V$}{
	Guess $s_{\ell} \in Q$ and store it\;
	\lIf{$s_{\ell}$ is not reachable from $i$}{do an infinite loop}
}
\ForEach{$(\ell,x,m) \in \circ$}{
	Store $(s_{\ell},x,s_{m})$\;
}
\While{there exists a stored tuple $(s_{\ell},x,s_{m})$ such that $s_{\ell} \neq s_{m}$}{
  Guess $a \in A$\;
  Guess $y \in X$\;
  \ForEach{stored tuple $(s_{\ell'},z,s_{m'})$ such that $y \preceq z$}{
	$(s_{\ell'},z,s_{m'}) \leftarrow (s_{\ell'} \cdot a,z,s_{m'})$
	}
}
\caption{Detecting the Subword-Pattern $\mathcal{P} =  \left( \mathcal{S},j \not \leq k \right)$ where $\mathcal{S} = \left( V,X,\circ \right)$}
\label{alg:subwordpattern}
\end{algorithm}

The following Lemma shows that the algorithm indeed finds the desired patterns. 

\begin{lemma}\label{lem:AlgorithmDoesWhatItIsSupposedToDo}
	A pattern $\mathcal{P}$ with stable superwords is present in $\mathcal{A}$ if and only if Algorithm \ref{alg:subwordpattern} terminates and at the end the stored states $s_j$, $s_k$ satisfy $s_j \not \semleq{\mathcal{A}} s_k$.
\end{lemma}

\begin{proof}
	Suppose first that $\mathcal{P}$ is present in $\mathcal{A}$ where $h$ and $g$ has the desired properties. We guess $s_{\ell} = g(\ell)$ in the first for-loop.

	For each variable $x$ and each step $n$ of the algorithm we have a homomorphism $q_n: X^* \to A^*$ with the following invariant properties:
	\begin{enumerate}[\itshape(i)]
		\item For each stored tuple $(s_{\ell}, x, s_{\ell'})$, we have $s_{\ell} \cdot q_n(x) = s_{\ell'}$,\label{aaa:invariant}
		\item if $x \preceq y$, then $\alp(q_n(x)) \subseteq \alp(q_n(y))$.\label{bbb:invariant}
		\item for all $x$, we have $\alp(q_{n}(x)) \subseteq \alp(h(x))$.\label{ccc:invariant}
	\end{enumerate}
	Note that if $q_n(x) = \varepsilon$ for all $x$, then the algorithm halts. Furthermore, if the algorithm halts, we have $s_j = g(j) \not\semleq{\mathcal{A}} g(k) = s_k$, which is the desired criteria. Thus, we need only show that the algorithm halts. We choose $q_1 = h$.

	Suppose there is some $q_n(x)$ which is nonempty. We show that it is possible to find $n' \geq n$ such that the above invariants are satisfied and $q_n(y) = q_{n'}(y)$ for all $y \not \succ x$ while either $|q_{n'}(x)| < |q_n(x)|$ or there is some $z \prec x$ such that $q_{n'}(z)$ is nonempty. We proceed by induction over $\preceq$ reversed.

	By induction, we can assume that there is $n''$ such that $q_{n''}(x)[1] = q_{n''}(y)[1] = a$ for all $y \succeq x$. If $x$ is maximal with respect to $\preceq$, it is trivially true. Otherwise, since $\alp(q_{n}(x)) \subseteq \alp(q_{n}(y))$ we use induction to remove letters from $q_n(y)$ until $a$ appears.

	For step $n' = n'' + 1$, we guess $a \in A$ and $x \in X$ in the interior of the while-loop. For each $y \succeq x$, let $t_y$ be such that $q_{n''}(y) = at_y$. We define
	\begin{align*}
		q_{n+1}(x) & =
		\begin{cases}
			t_x & \text{if $q_{n''}(z) = \varepsilon$ for all $z \prec x$}
			\\ t_xh(y) & \text{otherwise}
		\end{cases}
		\\ q_{n+1}(y) & =
		\begin{cases}
			t_y h(y) & \text{if $y \succ x$}
			\\ q_n(y) & \text{otherwise}
		\end{cases}
	\end{align*}
	We note that condition \itref{ccc:invariant} is satisfied.
	We also have that \itref{aaa:invariant} holds because $\mathcal{P}$ has stable superwords (if $y \succeq x$, then $y$ is a loop at $\ell'$, we must have $h(y)$ a cycle at $s_{\ell'}$ in the given tuple).

	To see that \itref{bbb:invariant} is satisfied, suppose $y \preceq z$. If $z \not \succeq x$ then $y \not \succeq x$ and thus the corresponding alphabets remain unchanged. Suppose instead $z \succeq x$. Then $\alp(q_{n+1}(y)) \subseteq \alp(h(y)) \subseteq \alp(h(z)) = \alp(q_{n+1}(z))$ giving the desired result.

	It now follows easily that we can make $q_n(x)$ empty for all $x$. Indeed, choose a nonempty $x$ which is minimal with respect to $\preceq$. By repeated application of the above argument, we can find $n'$ such that $q_{n'}(x) = \varepsilon$, while $q_n(y) = \varepsilon$ implies $q_{n'}(y) = \varepsilon$ (since if $q_n(y) = \varepsilon$, condition \itref{bbb:invariant} ensures that $x \not \preceq y$). We can thus make the $q_n(x)$ empty one by one.

	For the other direction, we need to define $g$ and $h$ with the desired properties. We define $g(\ell) = s_{\ell}$ for $\ell \in V$. Furthermore, for each $x \in X$, the algorithm provides a word $u_{x}$ being the concatenation of every $a$ guessed whenever $x$ was among the variables updated in an iteration of the while-loop. It is clear that if $x$ is an edge between $\ell$ and $\ell'$, then $s_{\ell} \cdot u_{x} = s_{\ell'}$ and if $x \preceq y$, then $u_{x}$ is a subword of  $u_{y}$. Thus, defining $h(x) = u_{x}$ gives the desired function. 
\end{proof}

\begin{proposition}\label{prp:ExistsNLAlgorithm}
	Let $\mathcal{A}$ be a DFA or a Carton--Michel automata. Checking the presence in $\mathcal{A}$ of a pattern $\mathcal{P}$ with stable superwords is in $\NL$ in the size of $\mathcal{A}$.
\end{proposition}

\begin{proof}
	For a fixed $\mathcal{P}$, Algorithm \ref{alg:subwordpattern} stores a fixed number of states in $Q$, and thus the algorithm is in $\NL$. It is a standard result that checking whether $s_j \not \semleq{\mathcal{A}} s_k$ is in $\NL$. Thus, the desired result follows from Lemma \ref{lem:AlgorithmDoesWhatItIsSupposedToDo}. 
\end{proof}

We note that Algorithm \ref{alg:subwordpattern} can be extended so that it checks whether an edge is final. Indeed, for each final edge $e$ of the pattern, we store a boolean $b_e$ which is set to true whenever the corresponding tuple encounters a final state. Similarly, we can check that $e$ maps to a nonempty word by storing a boolean which is set to true whenever the tuple is part of edges treated in the interior of the while loop. Thus, checking presence of enhanced subword-patterns is also in $\NL$.

We also give a hardness result. This is done via a reduction from graph reachability, a well known $\NL$-complete problem. This hardness result extends further than variety membership. Indeed, for DFAs we consider all (non-trivial) properties $P$ for which $L \in P$ implies $Lu^{-1} \in P$ and for Carton-Michel automata we consider all non-trivial properties for which $L \in P$ implies $u^{-1}L \in P$.
Note in particular that being in a language variety or being open/closed in the Cantor or alphabetic topology are properties with this trait.

\begin{proposition}\label{prp:NLHardness}
	Let $P$ be a nontrivial property of regular (resp. $\omega$-regular) languages containing the empty language and such that whenever $L \in P$ then $Lu^{-1} \in P$ (resp. $u^{-1}L \in P$). Given a DFA (resp. Carton-Michel automata) $\mathcal{A}$, deciding such a property is $\NL$-hard in the size of $\mathcal{A}$.
\end{proposition}

\begin{proof}
	We first consider the DFA case. Let $\mathcal{B} = \left( Q,A,\cdot,i,F \right)$ be a DFA such that $L(\mathcal{B}) \notin P$ (such a $\mathcal{B}$ exists since $P$ is nontrivial). Suppose we are given a digraph $\mathcal{G} = (V,E)$ and states $j,k$ where we want to check whether $k$ is reachable from $j$. Let $\#$ be an arbitrary symbol. We define the automata
	\begin{equation*}
		\mathcal{C} = \left( Q \cup V \cup \left\{ s \right\}, A \cup E \cup \left\{ \# \right\}, \circ, i, \left\{ k \right\} \right)
	\end{equation*}
	where $\circ$ is defined as
	\begin{equation*}
		\ell \circ x = \begin{cases}
			\ell \cdot x & \text{if $\ell \cdot x$ is defined}
			\\ \ell' & \text{if $x = (\ell,\ell') \in E$}
			\\ j & \text{if $x = \#$ and $\ell \in F$}
			\\ s & \text{otherwise}.
		\end{cases}
	\end{equation*}
	Intuitively, we have the following picture:
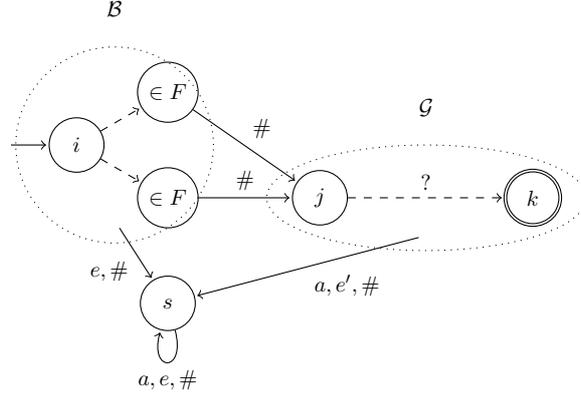
\begin{equation*}
	\begin{tikzpicture}[shorten >=1pt,node distance=1.4cm,on grid,auto,initial text = {},every node/.style={scale = 0.75}] 
	   \node[state] (q_0) {$j$}; 
	   \node (dist) [right=of q_0] {}; 
	   \node[state,accepting] (q_m1) [right=of dist] {$k$}; 
	   \draw[dotted] (dist) ellipse (2.1 and 0.7);
	   \node [above] (graph) [above=1.2 of dist] {$\mathcal{G}$};
	    \path[->] (q_0) edge[dashed]  node {?} (q_m1);
	    \node[state] (q_1) [left=2 of q_0] {$\in F$};
	    \node[state] (q_2) [above=of q_1] {$\in F$};
	    \node (q_4) [left=of q_2] {};
	    \node (el) at ($(q_1)!0.5!(q_4)$) {};
	    \node[state,initial] (q_3) [left=0.5 of el] {$i$};
	    \path[->] (q_3) edge[dashed] (q_2);
	    \path[->] (q_3) edge[dashed] (q_1);
	   \node [above] (aut) [above=1.8 of el] {$\mathcal{B}$};
	   \draw[dotted] (el) circle (1.3);
	   \path[->] (q_1) edge node {$\#$} (q_0);
	   \path[->] (q_2) edge node {$\#$} (q_0);
	   \node[state] (s) [below=of q_1] {$s$};
	   \node (aestart) [below=0.5 of dist] {};
	   \node (astart) [below=1 of el] {};
	   \path[->] (aestart) edge node[below right] {$a,e',\#$} (s);
	   \path[->] (astart) edge node[below left] {$e,\#$} (s);
	   \path[->] (s) edge[loop below]  node {$a,e,\#$} (s);
	\end{tikzpicture}
\end{equation*}
where we interpret $a$ as any letter in $A$, $e$ as any edge in $E$ and $e'$ as any edge in $E$ for which the transition is not already defined. It is clear that $\mathcal{C}$ is a DFA, and that its size is polynomial in the size of $\mathcal{G}$.

\begin{claim}
	The automaton $\mathcal{C}$ satisfies $P$ if and only if there is a path from $j$ to $k$.
\end{claim}

\begin{proof}
	Since $k$ is the only final state, it is clear that if there is no path from $j$ to $k$, then $L(\mathcal{C})$ is empty and thus satisfy $P$.
	On the other hand, suppose that there is a path labeled by $u$ from $j$ to $k$. We note that $L(\mathcal{B}) = L(\mathcal{C})(\#u)^{-1}$. Thus, if $L(\mathcal{C}) \in P$, then $L(\mathcal{B}) \in P$, a contradiction.
\end{proof}

Since $\mathcal{B}$ is fixed given a fixed property $P$, and the size of $\mathcal{C}$ is polynomial in the size of $\mathcal{G}$, we have a reduction from graph reachability to membership of $P$.

	The proof for Carton-Michel automata follows the same line of argument as in the DFA case.
	Let $\mathcal{B} = \left( Q,A,\cdot,I,F \right)$ be a Carton-Michel automata such that $L(\mathcal{B}) \notin P$, and suppose $\mathcal{G} = (V,E)$ and states $j,k$ are given such that we want to check whether $k$ is reachable from $j$. Let $\#$ again be an arbitrary symbol. We define the automata
	\begin{equation*}
		\mathcal{C} = \left( Q \cup V \cup \left\{ s, f, g \right\}, A \cup E \cup \left\{ \# \right\}, \circ, \left\{ j \right\}, F \cup \left\{ f \right\} \right)
	\end{equation*}
	where $\circ$ is defined as
	\begin{equation*}
		x \circ \ell = \begin{cases}
			x \cdot \ell & \text{if $x \cdot \ell$ is defined}
			\\ \ell' & \text{if $x = (\ell',\ell) \in E$}
			\\ k & \text{if $x = \#$ and $\ell \in I$}
			\\ f & \text{if $\ell \in \left\{ f,g \right\}$ and $x \in E \cup \left\{ \# \right\}$}.
			\\ g & \text{if $\ell \in \left\{ f,g \right\}$ and $x \in A$}
			\\ s & \text{otherwise}.
		\end{cases}
	\end{equation*}
	This gives the following picture. We note the similarity to the DFA case when changing the direction of all arrows.
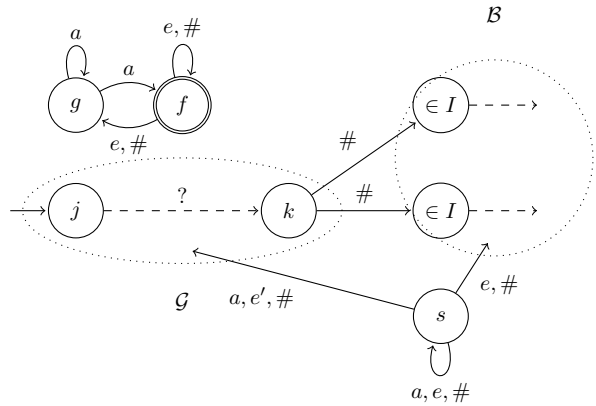
\begin{equation*}
	\begin{tikzpicture}[shorten >=1pt,node distance=1.4cm,on grid,auto,initial text = {},every node/.style={scale = 0.75}] 
	   \node[state] (q_0) {$k$}; 
	   \node (dist) [left=of q_0] {}; 
	   \node[state,initial] (q_m1) [left=of dist] {$j$}; 
	   \draw[dotted] (dist) ellipse (2.1 and 0.7);
	   \node [above] (graph) [below=1.2 of dist] {$\mathcal{G}$};
	    \path[->] (q_m1) edge[dashed]  node {?} (q_0);
	    \node[state] (q_1) [right=2 of q_0] {$\in I$};
	    \node[state] (q_2) [above=of q_1] {$\in I$};
	    \node (q_3) [right=of q_1] {};
	    \node (q_4) [right=of q_2] {};
	    \path[->] (q_1) edge[dashed] (q_3);
	    \path[->] (q_2) edge[dashed] (q_4);
	    \node (el) at ($(q_1)!0.5!(q_4)$) {};
	   \node [above] (aut) [above=1.9 of el] {$\mathcal{B}$};
	   \draw[dotted] (el) circle (1.3);
	   \path[->] (q_0) edge node {$\#$} (q_1);
	   \path[->] (q_0) edge node {$\#$} (q_2);
	   \node[state] (s) [below=of q_1] {$s$};
	   \node (aestart) [below=0.5 of dist] {};
	   \node (astart) [below=1 of el] {};
	   \path[->] (s) edge node[below left] {$a,e',\#$} (aestart);
	   \path[->] (s) edge node[below right] {$e,\#$} (astart);
	   \path[->] (s) edge[loop below]  node {$a,e,\#$} (s);
	   \node[state] (g) [above=of q_m1] {$g$};
	   \node[state,accepting] (f) [right=of g] {$f$};
	   \path[->] (g) edge[bend left]  node {$a$} (f);
	   \path[->] (f) edge[bend left]  node {$e,\#$} (g);
	   \path[->] (g) edge[loop above]  node {$a$} (g);
	   \path[->] (f) edge[loop above]  node {$e,\#$} (f);
	\end{tikzpicture}
\end{equation*}
We make the same interpretations of $a$, $e$ and $e'$ as in the DFA case.

\begin{claim}
	The automaton $\mathcal{C}$ is a Carton-Michel automaton.
\end{claim}

\begin{proof}\let\qed\relax
	We define $B = A \cup E \cup \left\{ \# \right\}$.
	Let $\alpha \in B^{\omega}$. We show that $\alpha$ has a unique final path in $\mathcal{C}$ by distinguishing two cases, either $\im(\alpha) \subseteq A$ or $\im(\alpha) \not\subseteq A$. Suppose $\im(\alpha) \not\subseteq A$. The only final loops containing letters outside $A$ is those contained in the component with $g$ and $f$, and it is clear that there is exactly one final path for each such word in that component.

	Next, suppose $\im(\alpha) \subseteq A$. In particular, we can write $\alpha = u \alpha'$ where $\alp(\alpha') = A$ and $u$ is either empty or ends with a letter which is not in $A$. Since $\mathcal{A}$ is a Carton Michel automata, there exists a unique run of $\alpha'$ in $\mathcal{A}$. Since the only added final state in $\mathcal{B}$ is $f$, and since any final state involving $f$ requires some $e \in E$ or $\#$ to appear infinitely often, the unique run of $\alpha'$ in $\mathcal{A}$ is also a unique run of $\alpha$ in $\mathcal{B}$. Let the start of this unique run be $\ell$. Since $\mathcal{B}$ is reverse deterministic, there exists a unique state in $\mathcal{B}$, say $\ell'$ such that $u \circ \ell = \ell'$. Hence, there is a unique run of $\alpha$ starting at $\ell'$.
\end{proof}

It is straightforward to generalise the previous claim to the Carton-Michel automata case; the automaton $\mathcal{C}$ satisfy $P$ if and only if there is a path from $j$ to $k$. Hence, we again have a reduction from graph reachability, giving the desired result. 
\end{proof}

\section*{Conclusion}

For all full and half levels of the $\FO^2$ quantifier alternation hierarchy, we give automata characterizations in terms of forbidden subword-patterns. These results rely on algebraic and topological characterizations of the $\FO^2$ levels (see Table~\ref{tbl:MonoidCriteria}). For finite words, we consider DFAs (Corollary~\ref{cor:PatternsForLevels}) and for infinite words, our patterns apply to Carton-Michel automata (Theorem~\ref{thm:Characterization} and Proposition~\ref{prp:CantorOpenCanBeCheckedByPattern}). For every fixed level, these patterns yield an $\NL$-algorithm to decide whether a given automaton accepts a language at this level (Proposition~\ref{prp:ExistsNLAlgorithm}); this problem is sometimes called the membership problem for the respective level. Together with a more general $\NL$-hardness result (Proposition~\ref{prp:NLHardness}), this shows that the membership problem is $\NL$-complete for every level of the $\FO^2$ quantifier alternation hierarchy for both finite and  infinite words.

\bibliographystyle{abbrv}
\bibliography{library}

\pagebreak

\appendix

\end{document}